\documentclass[10pt,journal,compsoc]{IEEEtran}
\ifCLASSOPTIONcompsoc
  \usepackage[nocompress]{cite}
\else
  \usepackage{cite}
\fi
\ifCLASSINFOpdf
\else
\fi
\usepackage{booktabs} 
\usepackage{tabularx}
\usepackage{graphicx}
\usepackage{url}
\usepackage{mdwlist}
\usepackage{subcaption}
\usepackage{hyperref}
\hypersetup{colorlinks=false,linkcolor=blue,urlcolor=blue,citecolor=red}
\usepackage{amsfonts}
\usepackage[labelfont=bf,textfont=bf]{caption}
\usepackage{multirow}
\usepackage{array}
\usepackage{xcolor}
\usepackage[flushleft]{threeparttable}
\usepackage{enumitem}

\usepackage{mathrsfs}
\usepackage{bm}
\usepackage[normalem]{ulem}
\useunder{\uline}{\ul}{}
\usepackage{tabu}
\usepackage{makecell}

\usepackage{caption}
\usepackage{float}
\usepackage[linesnumbered,ruled,commentsnumbered,vlined]{algorithm2e}
\usepackage{epstopdf}
\usepackage{color}
\usepackage[T1]{fontenc}
\usepackage{aecompl}
\usepackage{amsmath,amssymb,amsthm}

\newcommand{\ie}{\emph{i.e., }}

\newtheorem{theorem}{Theorem}
\newtheorem{lemma}{Lemma}
\begin{document}


\title{Learning Vertex Representations for \\Bipartite Networks}
\author{Ming~Gao, Xiangnan~He, Leihui~Chen, Tingting~Liu, Jinglin~Zhang$^{*}$ and Aoying~Zhou$^{*}$ \thanks{This paper is an extended version of the SIGIR '18 conference paper~\cite{DBLP:conf/sigir/0001C0Z18}. Jinglin~Zhang and Aoying~Zhou are the corresponding authors.}
\IEEEcompsocitemizethanks{
	\IEEEcompsocthanksitem Ming Gao is with the School of Data Science and Engineering, and KLATASDS-MOE in the School of Statistics, East China Normal University, Shanghai, China, 200062. mgao@dase.ecnu.edu.cn \protect\\
	
	\IEEEcompsocthanksitem Leihui Chen, Tingting Liu and Aoying Zhou are with the School of Data Science and Engineering, East China Normal University, Shanghai, China, 200062. leihuichen@gmail.com, tingtingliu223@gmail.com, ayzhou@dase.ecnu.edu.cn\protect\\
	
	\IEEEcompsocthanksitem Xiangnan He is with the School of Information Science and Technology, University of Science and Technology of China, Hefei, China. xiangnanhe@gmail.com. \protect\\
	
	\IEEEcompsocthanksitem Jinglin Zhang is with the Key Laboratory of Meteorological Disaster, Ministry of Education, Nanjing University of Information Science and Technology, Nanjing, China. Jinglin.zhang@nuist.edu.cn. \protect\\
}
}

\markboth{IEEE Transactions on Knowledge and Data Engineering, Submission 2018}
{}

\IEEEtitleabstractindextext{
\begin{abstract}
Recent years have witnessed a widespread increase of interest in network representation learning (NRL). By far most research efforts have focused on NRL for homogeneous networks like social networks where vertices are of the same type, or heterogeneous networks like knowledge graphs where vertices (and/or edges) are of different types. There has been relatively little research dedicated to NRL for bipartite networks. Arguably, generic network embedding methods like node2vec and LINE can also be applied to learn vertex embeddings for bipartite networks by ignoring the vertex type information. However, these methods are suboptimal in doing so, since real-world bipartite networks concern the relationship between two types of entities, which usually exhibit different properties and patterns from other types of network data. For example, E-Commerce recommender systems need to capture the collaborative filtering patterns between customers and products, and search engines need to consider the matching signals between queries and webpages. 

This work addresses the research gap of learning vertex representations for bipartite networks. We present a new solution BiNE, short for \textit{\textbf{Bi}partite \textbf{N}etwork \textbf{E}mbedding}, which accounts for two special properties of bipartite networks: long-tail distribution of vertex degrees and implicit connectivity relations between vertices of the same type. Technically speaking, we make three contributions: (1) We design a biased random walk generator to generate vertex sequences that preserve the long-tail distribution of vertices; (2) We propose a new optimization framework by simultaneously modeling the explicit relations (\ie observed links) and implicit relations (\ie unobserved but transitive links); (3) We explore the theoretical foundations of BiNE to shed light on how it works, proving that BiNE can be interpreted as factorizing multiple matrices. 
We perform extensive experiments on five real datasets covering the tasks of link prediction (classification) and recommendation (ranking), empirically verifying the effectiveness and rationality of BiNE. Our experiment codes are available at: \url{https://github.com/clhchtcjj/BiNE}. 

\end{abstract}
}

\maketitle
\IEEEdisplaynontitleabstractindextext
\IEEEpeerreviewmaketitle


\IEEEraisesectionheading{\section{Introduction}
\label{sec:introduction}}

\IEEEPARstart{N}{etwork} provides a ubiquitous data structure to model interactions (\ie edges) among entities (\ie vertices), having been widely used in many applications such as social networks~\cite{silkroad}, knowledge graphs~\cite{DBLP:acl/CaoHJCL17}, recommender systems~\cite{TriRank}, among others~\cite{POH}. 
However, performing predictive analytics or mining knowledge directly on networks exhibits several challenges, such as  high computation complexity, low parallelizability, and inapplicability of machine learning methods~\cite{DBLP:journals/corr/abs-1711-08752}. To handle these challenges, substantial works focused on network representation learning (NRL). In particular, they represent a vertex as a learnable embedding vector, where the proximity between vectors encodes the information about network structure. Thus the vertex embedding can be fed into machines learning methods to address various task such as ranking, link prediction, clustering, visualization and so on.

Recent advances in NRL have primarily focused on homogeneous network like social networks where vertices having the same type~\cite{DBLP:conf/kdd/ZhuCW018,DBLP:conf/kdd/PerozziAS14,DBLP:journals/corr/LiaoHZC17,DBLP:conf/cikm/LiDHTCL17}, or heterogeneous networks like knowledge graphs where vertices (and/or edges) are of different types~\cite{dong2017metapath2vec,DBLP:conf/kdd/ChangHTQAH15,DBLP:conf/wsdm/XuWCY17}. The effectiveness and prevalence of DeepWalk~\cite{DBLP:conf/kdd/PerozziAS14} inspire many works\cite{DBLP:conf/kdd/PerozziAS14,DBLP:conf/kdd/GroverL16} typically apply a two-step solution: 1) performing random walks, such as based random walk~\cite{DBLP:conf/kdd/GroverL16} or meta-path-based random walk~\cite{dong2017metapath2vec}, on the network to obtain a ``corpus'' of vertices; 2) applying embedding methods  on the corpus such as word2vec~\cite{DBLP:conf/nips/MikolovSCCD13} to obtain the embeddings for vertices.
     

However, there has been relatively little research dedicated to NRL for bipartite networks, for which there are two types of vertices~\cite{DBLP:journals/tkde/HeGKW17}. While generic network embedding methods like DeepWalk and node2vec can be applied to learn vertex representation for bipartite networks by ignoring the vertex type information, we argue that these methods are suboptimal in doing so because of two reasons. 
(1) Real-world bipartite networks concern the relationship between two types of entities, which usually exhibit different properties and patterns from other types of network data. For example, E-Commerce recommender system models the collaborative filtering patterns between customers and products, and Web search engine considers the matching signals between queries and webpages.
(2) The generated corpus may not preserve the characteristics of the bipartite network.
The power-law distribution is a common characteristic of many real-world bipartite networks~\cite{pongnumkul2018bipartite}, but the corpus generated by a universal random walk algorithm may not preserve this property, such as the one used in DeepWalk \cite{DBLP:conf/kdd/PerozziAS14}.
Specifically, it generates the same number of random walks starting from each vertices and constrains the length of random walks to be the same, which may limit the information of vertices with high degree and oversample vertices with low degree.

To address the limitations of existing methods on embedding bipartite networks, we devise a new solution for learning vertex representation in bipartite networks, namely BiNE (short for \textit{\textbf{Bi}partite \textbf{N}etwork \textbf{E}mbedding}).
%
We summarize the main contributions of this work as follows. 
\begin{enumerate}[leftmargin=*]
\item We propose a biased and self-adaptive random walk generator to preserve the long-tail distribution of vertex in bipartite networks as much as possible.
Specifically, we set the number of random walks starting from each vertex based on its importance and allow each walk to be stopped in a probabilistic way instead of setting a uniform length for all random walks. 
\item We propose a joint optimization framework to model the explicit and implicit relations simultaneously. Specifically, we devise a dedicated objective function for each relation and optimize them by sharing the embedding vectors, where different relations reinforce each other and lead to better vertex embeddings. 
\item We reveal the theoretical foundation of BiNE, which can be interpreted as implicitly factorizing multiple matrices, offering a better understanding how BiNE works.

\item We perform extensive experiments on several real datasets covering the tasks of links prediction (classification), recommendation (personalized ranking), and visualization, to illustrate the effectiveness and rationality of BiNE.
\end{enumerate}
A preliminary version of this work has been published in the conference of SIGIR 2018~\cite{DBLP:conf/sigir/0001C0Z18}. We summarize the main changes as follows:
\begin{enumerate}[leftmargin=*]
\item Introduction (Section~\ref{sec:introduction}). We have reconstructed the abstract and introduction to highlight the motivations of the extended version.
\item Theoretical foundations (Section~\ref{sec:BiMF}). We explore the theoretical foundations of BiNE to shed light on how it works, proving that BiNE can be interpreted as factorizing multiple matrices.
\item Experiments (Section~\ref{sec:experiments}). We add experiments to verify the proof, showing that the factorization-based implementation can achieve the same level of performance as BiNE in Section~\ref{subsec:RQ2}. In addition, we explore the performance of BiNE with different negative sampling strategies  to justify our sampler design in  Section~\ref{subsec:RQ4}.
\end{enumerate}
The remainder of the paper is organized as follows. We first formulate the problem in Section~\ref{sec:problem definition}, before delving into details of the method in Section~\ref{sec:BiNE}.  We show the theoretical connections with factorization methods in Section~\ref{sec:BiMF} and perform empirical studies in Section~\ref{sec:experiments}.
We review related work in Section~\ref{sec:related work} before concluding the paper in Section~\ref{sec:conclusion}.

\vspace{-10pt}
\section{Problem Formulation}
\label{sec:problem definition}
We first give notations used in this paper,
and then formalize the bipartite network embedding problem to be addressed.
%

\noindent\textbf{Notations}.
Let $G=(U,V,E)$ be a bipartite network,
where $U$ and $V$ denote the set of the two types of vertices respectively,
and $E\subseteq U\times V$ defines the inter-set edges.
%
$u_i$ and $v_j$ denote the $i$-th and $j$-th vertices in $U$ and $V$, respectively,
where $i=1,2,...,|U|$ and $j=1,2,...,|V|$.
Each edge carries a non-negative weight $w_{ij}$, describing the strength between
the connected vertices $u_i$ and $v_j$; if $u_i$ and $v_j$ are not connected, the edge weight $w_{ij}$ is set to zero.
Therefore, we can use a $|U| \times |V|$ matrix $\textbf{W}=[w_{ij}]$ to represent the weighted structure of the bipartite network.

\noindent\textbf{Problem Definition}.
The aim of bipartite network embedding is to map all vertices in the network into a low-dimensional embedding
space, where each vertex is represented as a dense vector.
In the embedding space, both the implicit relations between vertices of the same type and the explicit relations between vertices of different types should be preserved.
Formally, the problem can be defined as:
\begin{description}
	\item[\textbf{Input}:]A bipartite network $G=(U,V,E)$ and its weight matrix $\textbf{W}$.
	\item[\textbf{Output}:]A mapping function $f: U \cup V \rightarrow \mathbb{R}^{d}$, which maps
    each vertex in $G$ to a $d$-dimensional embedding vector.
\end{description}
To keep the notations simple, we use \textbf{${\textbf{u}_i}$} and \textbf{${\textbf{v}_j}$} to denote the
embedding vectors for vertices $u_i$ and $v_j$, respectively.
As such, we can present the embedding vectors of all vertices in the bipartite network as
two matrices $\textbf{U}=[{\textbf{u}_i}]$ and $\textbf{V}=[{\textbf{v}_j}]$.

The notations used in this paper are summarized in Table~\ref{tab:notaions and explanations}.

\begin{table}[htbp]
\centering
\caption{Notations used in this paper.}
\label{tab:notaions and explanations}
\begin{tabular}{|l|l|}
\hline
\multicolumn{1}{|c|}{\textbf{Notation}} & \multicolumn{1}{c|}{\textbf{Explanation}}   \\ \hline\hline
$G(U,V,E)$     & bipartite network  \\ \hline
$U=\{u_i\}_{i=1}^{|U|}$ &  the set of users    \\ \hline
$V=\{v_i\}_{i=1}^{|V|}$ &  the set of items    \\ \hline
$E\subseteq U\times V$ &  inter-set edges          \\ \hline
$\textbf{W}$ &  weight matrix \\ \hline
$w_{ij}$ &  an entry of weight matrix \textbf{W}\\ \hline
${\textbf{u}_i}$ & embedding vector of $u_i$    \\ \hline
${\textbf{v}_j}$ & embedding vector of $v_j$    \\ \hline
${\pmb{\theta}_i}$ & embedding vector of $u_i$ as context  \\ \hline
${\pmb{\vartheta}_j}$ &  embedding vector of $v_j$ as context  \\ \hline
$\textbf{U}=[{\textbf{u}_i}]$ &  user embedding vector    \\ \hline
$\textbf{V}=[{\textbf{v}_j}]$ &  item embedding vector    \\ \hline
$D^{(U)}$ &  corpus generated from vertex set $U$   \\ \hline
$D^{(V)}$ &  corpus generated from vertex set $V$   \\ \hline
$\#(u_i, u_j)$ & \# center-context vertex pairs in $D^{(U)}$  \\ \hline
$\#(v_i, v_j)$ & \# center-context vertex pairs in $D^{(V)}$  \\ \hline
\end{tabular}
\end{table}

\section{BiNE: Bipartite Network Embedding}
\label{sec:BiNE}
The typical objective in learning vertex embeddings is to be capable of reconstructing the network structure well~\cite{DBLP:conf/kdd/PerozziAS14,DBLP:conf/kdd/GroverL16}. While the structure of normal networks is mostly reflected in the observed edges, the case is more complicated for bipartite networks --- two vertices of the same type are not directly connected, but it doesn't necessarily mean that they do not have relation. This poses challenges to bipartite network embedding, such that modeling only observed edges is insufficient to retain the fidelity. Towards this end, we propose to account for both the observed edges (Section~\ref{subsec:Modeling Explicit Relations} Modeling Explicit Relations) and the unobserved but transitive edges (Section~\ref{subsec:Modeling Explicit Relations} Modeling Implicit Relations). The final vertex embeddings are achieved by jointly optimizing the two tasks (Secion~\ref{subsection: model joint optimization} Joint Optimization). 

\subsection{Modeling Explicit Relations}
\label{subsec:Modeling Explicit Relations}
Edges between vertices of different types in a bipartite afford us a signal to capture the explicit structure information.
Similar to the modeling of 1st-order proximity in LINE~\cite{DBLP:conf/www/TangQWZYM15}, we preserve the explicit structure information by minimizing the difference between the empirical distribution of vertex co-occurring probability and the reconstructed distribution by the vertex embeddings.
The co-occurring probability between two connected vertices $u_i$ and $v_j$ in the original bipartite network is defined as:
\vspace{-5pt}
\begin{equation}\small
\begin{aligned}
\label{eq:empirical pro}
P(i,j)=\frac{w_{ij}}{\sum_{e_{st}\in E}w_{st}}.
\end{aligned}
\vspace{-5pt}
\end{equation}
where $w_{ij}$ is the weight of edge $e_{ij}$.
%
In addition, the local proximity between them in the embedding space can be estimated by their inner product~\cite{DBLP:conf/www/TangQWZYM15,DBLP:conf/kdd/PerozziAS14,DBLP:conf/kdd/GroverL16}, we further transform this interaction value to the probability space by the sigmoid function:
\vspace{-5pt}
\begin{equation}\small
\begin{aligned}
\label{eq:joint pro}
\hat P(i,j)=\frac{1}{1+exp(-{\textbf{u}_i}^T{\textbf{v}_j})}.
\end{aligned}
\vspace{-5pt}
\end{equation}
where ${\textbf{u}_i} \in \mathbb{R}^{d}$ and ${\textbf{v}_j} \in \mathbb{R}^{d}$ are the embedding vectors of vertices $u_i$ and $v_j$, respectively.

After getting the empirical distribution and the reconstructed distribution, we employ the KL-divergence to measure the difference between  the two distributions, and learn the embedding vectors by minimizing the difference.
Thus the objective function can be defined as:
\vspace{-5pt}
\begin{equation}\small
\begin{aligned}
\label{eq:O_1}
minimize \quad O_1 &= KL(P||\hat P)
                   = \sum_{e_{ij}\in E} P(i,j)\log(\frac{P(i,j)}{\hat P(i,j)}) \\
                   &\propto -\sum_{e_{ij}\in E} w_{ij}\log \hat P(i,j).
\end{aligned}
\vspace{-5pt}
\end{equation}
Intuitively, two strongly connected vertices in the original network will be close with each other in the embedding space by minimizing the objective function.
Thus the explicit structure information can be preserved.

\subsection{Modeling Implicit Relations}
The effectiveness of modeling implicit relations in recommendation~\cite{DBLP:conf/AAAI/LuYu,DBLP:conf/aaai/JiangCYXY16} (which deals with user-item bipartite network) motivates us to explore the implicit relations in bipartite networks towards real-world applications.
Although a perfect reconstruction of explicit relations can fully recover the implicit relations, it is impractical to rely on this. As such, we speculate that modeling the implicit relations between vertices of the same type could bring extra benefits to explicit relation modeling. 
Intuitively, if there exists a path between two vertices, it implies certain implicit relation between them; the number of the paths and their length indicate the strength of the relation. However, counting the paths between two vertices comes at the cost of very high complexity, which is unaffordable for large-scale networks. To encode such high-order implicit relations among vertices in a bipartite network, we resort to the solution of DeepWalk. To be exact, the bipartite network is first converted to two corpora of vertex sequences by performing random walks; then the embeddings are learned from the corpora which encodes high-order relations between vertices.


\subsubsection{\textbf{Constructing Corpus of Vertex Sequences}}
It is a common way to convert a network into a corpus of vertex sequences by performing random walks on the network, which has been used in some homogeneous network embedding methods~\cite{DBLP:conf/kdd/PerozziAS14,DBLP:conf/kdd/GroverL16}.
However, directly performing random walks on a bipartite network could fail, since there is no stationary distribution of random walks on bipartite networks due to the periodicity issue~\cite{DBLP:conf/complenet/AlzahraniHB14}.
To address this issue, we consider performing random walks on two homogeneous networks that contain the 2nd-order proximity between vertices of the same type.
Following the idea of Co-HITS~\cite{DBLP:conf/kdd/DengLK09}, we define the 2nd-order proximity between two vertices as:
\vspace{-1pt}
\begin{equation}\small
\begin{aligned}
\label{eq:one side vertex proximity}
w^{U}_{ij} = \sum_{k\in V}w_{ik}w_{jk}; \;\;
w^{V}_{ij} = \sum_{k\in U}w_{ki}w_{kj}.
\end{aligned}
\vspace{-1pt}
\end{equation}
where $w_{ij}$ is the weight of edge $e_{ij}$.
Hence, we can use the $|U| \times |U|$ matrix $\textbf{W}^{U}=[w^{U}_{ij}]$ and the $|V| \times |V|$ matrix $\textbf{W}^{V}=[w^{V}_{ij}]$
to represent the two induced homogeneous networks, respectively.
%
%

Now we can perform truncated random walks on the two homogeneous networks to generate two corpora for learning the high-order implicit relations.
%
To generate a corpus with a high fidelity, we propose a biased and self-adaptive random walk generator, which can preserve the vertex distribution in a bipartite network.
We highlight its core designs as follows:
\begin{itemize}[leftmargin=*]
\item First, we relate the number of random walks starting from each vertex to be dependent on its importance, which can be measured by its centrality.
%
For a vertex, the greater its centrality is, the more likely a random walk will start from it.
As a result, the vertex importance can be preserved to some extent.

\item We assign a probability to stop a random walk in each step.
In contrast to DeepWalk and other work~\cite{dong2017metapath2vec} that apply a fixed length on the random walk, we allow the generated vertex sequences have a variable length, in order to have a close analogy to the variable-length sentences in natural languages.
\end{itemize}
Generally speaking, the above generation process follows the principle of ``rich gets richer'', which is a physical phenomena existing in many real networks, i.e., the vertex connectivities follow a scale-free power-law distribution~\cite{DBLP:journals/jmlr/LeskovecCKFG10}. 

The workflow of our random walk generator is summarized  in Algorithm~\ref{alg:alg1},
where $maxT$ and $minT$ are the maximal and minimal numbers of random walks starting from all vertices, respectively.
$\mathcal{D}^U$ (or $\mathcal{D}^V$) output by Algorithm~\ref{alg:alg1} is the corpus generated from the vertex set $U$ (or $V$).
The vertex centrality can be measured by many metrics, such as degree centrality, PageRank
and HITS~\cite{DBLP:conf/soda/Kleinberg98}, etc.,
and we use HITS in our experiments.
$BiasedRandomWalk(W^{R} ,v_i, p)$ generates a vertex sequence, which starts from vertex $v_i$.  
For determining whether a vertex is the destination of the sequence or not, it will generate a uniform random variable from $U[0,1]$.
The sequence will terminate if the generated uniform r.v. is less than the stopping probability $p$.
Thus, the length of the generated vertex sequence is controlled by the walk stopping probability.
%

\IncMargin{0.5em}
 \begin{algorithm}
  \SetKwData{Left}{left}\SetKwData{This}{this}
  \SetKwData{Up}{up}\SetKwFunction{Union}{Union}
  \SetKwFunction{FindCompress}{FindCompress}\SetKwInOut{Input}{Input}\SetKwInOut{Output}{Output}
  \Input{weight matrix of the bipartite network $\textbf{W}$,
        vertex set $R$ (can be $U$ or $V$),
        maximal walks per vertex $maxT$,
        minimal walks per vertex $minT$,
        walk stopping probability $p$\\
  }
  \Output{a set of vertex sequences $\mathcal{D}^R$\\
  }
  \BlankLine
  Calculate vertices' centrality: $\textbf{H} = CentralityMeasure(\textbf{W})$\;
  Calculate $W^{R}$ w.r.t. Equation (\ref{eq:one side vertex proximity})\;
  \ForEach{vertex $v_i\in R$}{
    $l = \max (\textbf{H}(v_i) \times maxT, minT)$\;
    \For{$i = 0$ to $l$}{
        $\mathcal{D}_{v_i} = BiasedRandomWalk(W^{R} ,v_i,p)$\;
        Add $\mathcal{D}_{v_i}$ into $\mathcal{D}^R$\;
    }
 }
 \Return $\mathcal{D}^R$\;
\caption{WalkGenerator($W$, $R$, $maxT$, $minT$, $p$)}
\label{alg:alg1}
\end{algorithm}
\DecMargin{0.5em}

\subsubsection{\textbf{Implicit Relation Modeling}}
After performing biased random walks on the two homogeneous networks respectively, we obtain two corpora of vertex sequences.  
Next we employ the Skip-gram model~\cite{DBLP:conf/nips/MikolovSCCD13} on
the two corpora to learn vertex embeddings. The aim is to capture the high-order proximity, which assumes that vertices frequently co-occurred in the same context of a sequence should be assigned to similar embeddings. 
Given a vertex sequence $S$ and a vertex $u_i$, the context is defined as the $ws$ vertices before $u_i$ and after $u_i$ in $S$; 
each vertex is associated with a context vector ${\pmb{\theta}_i}$ (or ${\pmb{\vartheta}_j}$) to denote its role as a context.
As there are two types of vertices in a bipartite network, we preserve the high-order proximities separately. Specifically, for the corpus $\mathcal{D}^U$, the conditional probability to maximize is:
\begin{eqnarray}\small
\begin{aligned}
\vspace{-2pt}
\label{eq:con1}
maximize \; O_2 &=& \prod_{u_i\in S\wedge S\in \mathcal{D}^U}\prod_{\substack{u_c\in C_S(u_i)}}P(u_c|u_i).\\
maximize \; O_3 &=& \prod_{v_j\in S\wedge S\in \mathcal{D}^V}\prod_{\substack{v_c\in C_S(v_j)}}P(v_c|v_j).
\end{aligned}
\vspace{-2pt}
\end{eqnarray}
where $C_S(u_i)$ (or $C_S(v_j)$) denotes the context vertices of vertex $u_i$ (or $v_j$) in a sequence $S$.

Following existing neural embedding methods~\cite{DBLP:conf/kdd/GroverL16,DBLP:conf/kdd/PerozziAS14,DBLP:conf/www/TangQWZYM15}, we parameterize the conditional probability $P(u_c|u_i)$ and $P(v_c|v_j)$ using the inner product kernel with softmax for output:
\begin{equation}\label{eq:cond}\small
\vspace{-2pt}
P(u_c|u_i) =  \frac{\exp{({\textbf{u}_i}^T{\pmb{\theta}_c})}}{\sum_{k=1}^{|U|}{\exp{({\textbf{u}_i}^T{\pmb{\theta}_k})}}}, \;\;
P(v_c|v_j) = \frac{\exp{({\textbf{v}_j}^T{\pmb{\vartheta}_c})}}{\sum_{k=1}^{|V|}{\exp{({\textbf{v}_j}^T{\pmb{\vartheta}_k})}}}.
\vspace{-1pt}
\end{equation}
where $P(u_c|u_i)$ denotes how likely $u_c$ is observed in the contexts of $u_i$; similar meaning applies to $P(v_c|v_j)$. 
With this definition, achieving the goal defined in Equations (\ref{eq:con1}) will force the vertices with the similar contexts to be close in the embedding space. 
Nevertheless, optimizing the objectives is non-trivial, since each evaluation of the softmax function needs to traverse all vertices of a side, which is very time-costing. 
To reduce the learning complexity, we employ the idea of negative sampling~\cite{DBLP:conf/nips/MikolovSCCD13}.

\subsubsection{\textbf{Negative Sampling}}
\label{subsubsec: negative sampling}
The idea of negative sampling is to approximate the costly denominator term of softmax with some sampled negative instances~\cite{DBLP:conf/icde/HongzhiYin}. Then the learning can be performed by optimizing a point-wise classification loss. 
For a center vertex $u_i$, high-quality negatives should be the vertices that are dissimilar from $u_i$. 
Towards this goal, some heuristics have been applied, such as frequency-based negative sampling method proposed to learn the representations for words~\cite{DBLP:conf/nips/MikolovSCCD13}. 
Specifically, words with high frequency have large probability of being chosen as negative instances, which is suitable for language model since high frequency words are useless words such as \emph{he, she, it, is, the}, etc.
Nevertheless, high frequency vertices in bipartite networks are often the most important entities, such as popular items or active users, and the tracing phenomenon widely exiting in user buying and watching behaviors indicates us that it might be suboptimal to simply treat the high frequency vertices as negative instances.
Here we propose a more grounded sampling method that caters the network data.  
%

First we employ locality sensitive hashing (LSH)~\cite{DBLP:conf/cikm/WangCSR13} to block vertices after shingling each vertex by its $ws$-hop neighbors with respect to the topological structure in the input bipartite network. 
Given a center vertex, we then randomly choose the negative samples from the buckets that are different from the bucket contained the center vertex. 
Through this way, we can obtain high-quality and diverse negative samples, since 
LSH can guarantee that dissimilar vertices are located in different buckets in a probabilistic way~\cite{DBLP:conf/cikm/WangCSR13}.

%
%
Let $N_S^{ns}(u_i)$ denote the $ns$ negative samples for a center vertex $u_i$ in sequence $S\in \mathcal{D}^U$, we can then approximate the conditional probability $p(u_c|u_i)$ defined in Equation~(\ref{eq:cond}) as:
\begin{equation}\small
p(u_c, N_{S}^{ns}(u_i)|u_i) = \prod_{\substack{z\in \{u_c\}\cup N_{S}^{ns}(u_i)}}P(z|u_i),\label{eq:n_2}
\end{equation}
where the probability $P(z|u_i)$ is defined as:
\begin{eqnarray*} \small
P(z|u_i) & = & \left\{
                      \begin{array}{ll}
                       \sigma({\textbf{u}_i}^T{\pmb{\theta}_z}) , & \hbox{if }z \hbox{ is a context of }u_i \\
                        1 - \sigma({\textbf{u}_i}^T{\pmb{\theta}_z}), & z\in N_{S}^{ns}(u_i)
                      \end{array}\right.,\label{eq:o_2}
\end{eqnarray*}
where $\sigma$ denotes the sigmoid function $1/(1+e^{-x})$. By replacing $p(u_c|u_i)$ in Equation~(\ref{eq:con1}) with the definition of $p(u_c, N_{S}^{ns}(u_i)|u_i)$, we can get the approximated objective function to optimize. 
The semantics is that the proximity between the center vertex and its contextual vertices should be maximized, whereas the proximity between the center vertex and the negative samples should be minimized. 

Following the similar formulations, we can get the counterparts for the conditional probability $p(v_c|v_j)$, the details of which are omitted here due to space limitation. 

\subsection{Joint Optimization}
\label{subsection: model joint optimization}
To embed a bipartite network by preserving both explicit and implicit relations simultaneously, we combine their objective functions to form a joint optimization framework.
%
\vspace{-1pt}
\begin{equation}
\vspace{-2pt}
\begin{aligned}
\label{eq:L}
maximize\; L = \alpha \log O_2 + \beta \log O_3 - \gamma O_1.
\end{aligned}
\vspace{-1pt}
\end{equation}
where parameters $\alpha$, $\beta$ and $\gamma$ are hyper-parameters to be specified to combine different components in the joint optimization framework.

%
%
%
To optimize the joint model, we utilize the Stochastic Gradient Ascent algorithm (SGA). 
Note that the three components of Equation (\ref{eq:L}) have different definitions of a training instance.
To handle this issue, we tweak the SGA algorithm by performing a gradient step as follows:
\\
\textbf{Step I:} For a stochastic explicit relation, i.e., an edge $e_{ij}\in E$, we first update the embedding vectors ${\textbf{u}_i}$ and ${\textbf{v}_j}$ by utilizing SGA to maximize the last component $L_1 = -\gamma O_1$. 
%
We give the SGA update rule for ${\textbf{u}_i}$ and ${\textbf{v}_j}$ as follows:
\vspace{-2pt}
\begin{eqnarray}\small
{\textbf{u}_i} &=& {\textbf{u}_i} + \lambda \{\gamma  w_{ij} [1-\sigma({\textbf{u}_i}^T{\textbf{v}_j})]\cdot {\textbf{v}_j}\},\label{eq:e_uv1} \nonumber \\
{\textbf{v}_j} &=& {\textbf{v}_j} + \lambda \{\gamma  w_{ij} [1-\sigma({\textbf{u}_i}^T{\textbf{v}_j})]\cdot {\textbf{u}_i}\}.\label{eq:e_uv2}
\vspace{-5pt}
\end{eqnarray}
\vspace{+2pt}
\textbf{Step II:} We then treat vertices $u_i$ and $v_j$ as center vertex; by employing SGA to maximize objective functions $L_2 = \alpha \log O_2$ and $L_3 = \beta \log O_3$, we can preserve the implicit relations.
Specifically, given the center vertex $u_i$ (or $v_j$) and its context vertex $u_c$ (or $v_c$),
we update their embedding vectors ${\textbf{u}_i}$ (or ${\textbf{v}_j}$) as follows:
%
\begin{eqnarray}\small
{\textbf{u}_i} & = & {\textbf{u}_i} + \lambda \{\sum_{\substack{z\in \{u_c\} \cup N_{S}^{ns}(u_i)}}\alpha [I(z, u_i)-\sigma({\textbf{u}_i}^T{\pmb{\theta}_z})]\cdot {\pmb{\theta}_z}\},\label{eq:e_uorv1}\nonumber\\
{\textbf{v}_j} & = & {\textbf{v}_j} + \lambda \{\sum_{\substack{z\in \{v_c\} \cup N_{S}^{ns}(v_j)}}\beta[I(z, v_j)-\sigma({\textbf{v}_j}^T{\pmb{\vartheta}_z})]\cdot {\pmb{\vartheta}_z}\}.\label{eq:e_uorv2}\nonumber\\
\vspace{-5pt}
\end{eqnarray}
where $I(z,u_i)$ is an indicator function that determines whether vertex $z$ is in the context of $u_i$ or not; similar meaning applies to $I(z,v_j)$.  
%
Furthermore, the context vectors of both positive and negative instances are updated as:
%
\begin{eqnarray}\small
{\pmb{\theta}_z} &=& {\pmb{\theta}_z} + \lambda \{\alpha [I(z, u_i)-\sigma({\textbf{u}_i}^T{\pmb{\theta}_z})]\cdot {\textbf{u}_i}\},\label{eq:p_uorv1} \nonumber \\
{\pmb{\vartheta}_z} &=& {\pmb{\vartheta}_z} + \lambda \{\beta [I(z, v_j)-\sigma({\textbf{v}_j}^T{\pmb{\vartheta}_z})]\cdot {\textbf{v}_j}\}.\label{eq:p_uorv2}
\vspace{-3pt}
\end{eqnarray}

We use the embedding vectors as the representations of vertices. Concatenating the embedding and contextual vectors for each vertex may improve the representations, which we leave as future work.
\subsection{Discussion}
%

\noindent\textbf{Computational Complexity Analysis}.
The corpus generation and joint model optimization are two key processes of BiNE.
Here we discuss the computational complexity of the two processes respectively.

For the large-scale network, the complexity of generating corpus will increase since $\textbf{W}^{U}$ and $\textbf{W}^{V}$ become large and dense.
To avoid processing the dense matrix, we directly perform two-step walk in the original bipartite network to generate corpora.
%
%
Let $vc$ denotes the visitation count of vertex $v$ in the generated corpus.
The context size is therefore $vc\cdot 2ws$, which is a big value for vertices having high degrees.
Yet we only randomly select a small batch, e.g., $bs$ ($bs \ll vc$), of the contextual                            vertices for each center vertex.
Thus, the complexity of algorithm is $O(2|E|\cdot bs\cdot 2ws \cdot(ns+1))$.
%
To some extent, all the contextual vertices of a center vertex can be trained in each iteration by setting a proper $bs$, because the center vertex will be visited more than once when traversing all edges. Consequently, the performance is also guaranteed while the executive efficiency of BiNE is greatly improved.

\section{BiNE as Factorizing Multiple Matrices}
\label{sec:BiMF}
Prior efforts have revealed that several network embedding methods like DeepWalk, LINE and node2vec can be understood as performing factorization on some purposefully designed matrices~\cite{DBLP:conf/wsdm/QiuDMLWT18}.
The fundamental reason is that these methods use inner product to measure the affinity of two vertices, which also forms the basis of matrix factorization~\cite{DBLP:conf/www/HeLZNHC17}. 
In this section, we prove that BiNE can been also understood as co-factorizing multiple purposefully designed matrices.
Thus, we propose a matrix factorization based embedding method, namely BiNE-MF, which is a two-step framework, including proximity matrix construction and matrix factorization.

%
\vspace{-10pt}
\subsection{Derivation of the Proximity Matrices}
\label{subsec:Characterizing Complement Matrix}
In the following analysis, let $\#(u_i,u_j)$ (or $\#(v_i,v_j)$) be the number of center-context vertex pairs in $\mathcal{D}^U$ (or $\mathcal{D}^V$). 
Moreover, $\#(u_i)=\sum_{j=1}^{|U|}\#(u_i,u_j)$ (or $\#(v_i)=\sum_{j=1}^{|V|}\#(v_i,v_j)$) denotes the frequency that vertex $u_i$ (or $v_i$) appeared in $\mathcal{D}^U$ (or $\mathcal{D}^V$).

\subsubsection{Global Objective Function Rewriting}
The optimization of $\log O_2$ in Equation~(\ref{eq:L}) is trained in an online fashion via stochastic gradient updates over the observed pairs $(u_i,u_j)$ in the corpus $\mathcal{D}^U$. 
Its global objective can be obtained by summing over the observed $(u_i,u_j)$ pairs in the corpus~\cite{DBLP:conf/nips/LevyG14}:
\begin{equation}
\begin{aligned}
\label{eq:L_2}
maximize\; \log O_2 &= \sum_{i=1}^{|U|}\sum_{j=1}^{|U|} \#(u_i,u_j)\cdot \ell_{uu}(i,j).
\end{aligned}
\end{equation}
where
\begin{equation}
\begin{aligned}
\label{eq:L_2(i,j)}
\ell_{uu}(i,j) = \log{\sigma({\textbf{u}_i}^T{\pmb{\theta}_j})}+\sum_{j'\in N_{S}^{ns}(u_i)}\log{\sigma(-{\textbf{u}_i}^T{\pmb{\theta}_{j'}})},
\end{aligned}
\end{equation}
which is the local objective function for a single center-context vertex pair $(u_i,u_j)$.

Let $x_{ij}^U \doteq {\textbf{u}_i}^T{\pmb{\theta}_j}$, then the local objective function $\ell_{uu}(i,j)$ can be treated as a function of $x_{ij}^U$.
Therefore, $\ell_{uu}(i,j)$ can be simplified as:
\vspace{-5pt}
\begin{equation}\small
\begin{aligned}
\label{eq:L_2(i,j)}
\ell_{uu}(i,j) = \log{\sigma(x_{ij}^U)}+\sum_{j'\in N_{S}^{ns}(u_i)}\log{\sigma(-x_{ij'}^U)}.
\end{aligned}
\vspace{-5pt}
\end{equation}
In BiNE, for each center vertex $u_i$, its negative sample $u_{j'}$ is uniformly sampled from $N_S^{ns}(u_i)$.
Thus, based on the importance sampling, the second term of Equation~(\ref{eq:L_2(i,j)}) can be approximated by the conditional expectation given $u_{j'}\in N_S^{ns}(u_i)$.
\vspace{-5pt}
\begin{equation}\small
\begin{aligned}
\label{eq:negative_item}
\sum_{j'\in N_{S}^{ns}(u_i)}\log{\sigma(-x_{ij'}^U)}
&=\sum_{j'\in N_{S}^{ns}(u_i)}\frac{\log{\sigma(-x_{ij'}^U)}}{p(j')}\cdot p(j')\\
&\approx E_{j'\sim p}\big[\frac{\log{\sigma(-x_{ij'}^U)}}{p(j')}|N_{S}^{ns}(u_i)\big]\\
&=ns\cdot E_p\big[\log{\sigma(-x_{ij'}^U)}|N_{S}^{ns}(u_i)\big].\\
\end{aligned}
\vspace{-5pt}
\end{equation}
where $p(j')=\frac{1}{ns}$ is a conditional probability when $u_{j'}$ samples from $N_{S}^{ns}(u_i)$.

Since $N_{S}^{ns}(u_i)$ is generated in a probabilistic manner, the conditional expectation is a random variable related to the center vertex $u_i$. 
As a result, the log-likelihood $\log O_2$ in Equation~(\ref{eq:L_2}) cannot be maximized directly. 
Therefore, we learn the maximum likelihood estimate (MLE) of parameters via applying the EM-algorithm.

\textbf{E-Step:} The expectation of $\ell_{uu}(i,j)$ is:
\vspace{-1pt}
\begin{equation}\small
\begin{aligned}
\label{eq:negative_item}
E[\ell_{uu}(i,j)] 
& = \log{\sigma(x_{ij}^U)}+ ns \cdot E\big[E_p\big[\log{\sigma(-x_{ij'}^U)}|N_{S}^{ns}(u_i)\big]\big]\\
& = \log{\sigma(x_{ij}^U)}+ ns \cdot E\big[\log{\sigma(-x_{ij'}^U)}\big].\\
\end{aligned}
\end{equation}
As mentioned in Section~\ref{subsubsec: negative sampling}, the LSH-based negative sample $u_{j'}$ is sampled from the buckets that are different from the bucket contained center vertex $u_i$.
Let $\mathcal{J}_{ij}$ be the Jaccard similarity between vertices $u_i$ and $u_j$, then the probability, that
$u_i$ and $u_j$ are mapped into the different buckets, can be computed as $q(u_i,u_j)=(1-(\mathcal{J}_{ij})^r)^b$, where $b$ and $r$ denote the number of bands and the number of rows in each band, which are two hyper-parameters in LSH.
It is noteworthy that the vertices are mapped into the same bucket of LSH with larger probability if they have a higher proximity, i.e., given center vertex $u_i$, a context vertex is mapped to the different buckets with a small probability.
Since both $q(u_i,u_j)$ and the number of context vertices are small, $E[\ell_{uu}(i,j)]$ can be approximated as:
\begin{equation}\small
\begin{aligned}
E[\ell_{uu}(i,j)] 
&= \log{\sigma(x_{ij}^U)} + ns\cdot \sum_{j'\in N_{S}(u_i)}q(u_i,u_{j'}) \cdot \log{\sigma(-x_{ij'}^U)}\\
&\approx \log{\sigma(x_{ij}^U)} + ns\cdot \sum_{j=1}^{|U|}q(u_i,u_j) \cdot \log{\sigma(-x_{ij}^U)}
\end{aligned}
\end{equation}
Naturally, the expectation of log-likelihood function $\log O_2$ can be approximated as:
\begin{equation}\small
\begin{aligned}
\label{eq:E_L_2}
&E[\log O_2]\\
&= \sum_{i,j} \#(u_i,u_j) \cdot E[\ell_{uu}(i,j)]\\
&= \sum_{i,j}\#(u_i,u_j)\cdot[\log{\sigma(x_{ij}^U)}
+ns\cdot \sum_{l=1}^{|U|}q(u_i,u_l) \cdot \log{\sigma(-x_{il}^U)}]\\
&= \sum_{i,j}[\#(u_i,u_j)\cdot\log{\sigma(x_{ij}^U)}
+ns\cdot\#(u_i)\cdot q(u_i,u_j)\cdot\log{\sigma(-x_{ij}^U)}].
\end{aligned}
\end{equation}

\textbf{M-Step:}
The optimal solution $(x_{ij}^U)^*$ can be obtained if we maximize $E[\log O_2]$ directly.
After taking the derivative of $E[\ell_{uu}]$ and setting $\frac{\partial{E[\ell_{uu}}]}{\partial{x_{ij}^U}} = 0$, we can obtain the MLE of $x_{ij}^U$:
\begin{equation}\small
\begin{aligned}
\label{eq:x_u}
(x_{ij}^U)^* =\log{\frac{\#(u_i,u_j)}{\#(u_i) \cdot q(u_i,u_j)}}-\log{ns},
\end{aligned}
\end{equation}
where $q(u_i, u_j)$ is the probability that $u_i$ and $u_j$ are mapped into the different buckets in the LSH, and 
\begin{equation}
\label{eq:conditional}
\frac{\#(u_i, u_j)}{\#(u_i)}
=\frac{\#(u_i, u_j)}{|\mathcal{D}|} \cdot \frac{|\mathcal{D}|}{\#(u_i)}
=\frac{p(u_i, u_j)}{p(u_i)}
=p(u_j|u_i)
\end{equation}
is the conditional probability, which can evaluate the proximity of center-context pair $(u_i, u_j)$ in the corpus $\mathcal{D}$. 
$(x_{ij}^U)^*$ tends to be a larger value if $u_i$ and $u_j$ are vertex pair with a higher proximity. 
This is due to the fact that $p(u_j|u_i)$ is larger and $q(u_i, u_j)$ is smaller in this case. 

Similarly, we can obtain the MLE of $(x_{ij}^V)$ via applying EM algorithm to $\log O_3$:
\begin{equation}\small
\begin{aligned}
\vspace{-2pt}
\label{eq:x_v}
(x_{ij}^V)^* =\log{\frac{\#(v_i,v_j)}{\#(v_i) \cdot q(v_i,v_j)}}-\log{ns}.
\end{aligned}
\end{equation}

\subsubsection{Approximating the Proximity Matrices}
Let $\textbf{M}^U_{ij} = (x_{ij}^U)^*$ be an entry of the proximity matrix $\textbf{M}^U$, which evaluates how similar between $u_i$ and $u_j$ in the embedding space. 
As shown in Equation~\ref{eq:x_u}, counter $\#(u_i,u_j)$ is dependent on the generated vertex sequences. 
Thus, proximity matrix $\textbf{M}^U$ is difficult to compute in practice.

To avoid generating the vertex sequences, we approximate the value of $\frac{\#(u_i,u_j)}{\#(u_i)}$ according to the following theories.

\begin{lemma}
\label{lemma:l1}
Let $\textbf{P}=\textbf{D}^{-1}\textbf{W}$ be the transition matrix of a random walk, where $\textbf{W}$ is the weighted matrix of a homogeneous network $G$; $\textbf{D}=diag(d_1,\cdots, d_i,\cdots)$, where $d_i$ represents the weighted degree of vertex $i$; $vol(G)=\sum_i\sum_j w_{ij}$; $L_\mathcal{D}$ is the length of corpus $\mathcal{D}$ generated by the random walk.
When $L_\mathcal{D} \rightarrow \infty$, $\frac{\#(u_i,u_j)}{|\mathcal{D}|}$ converges in probability $(\stackrel{p}{\rightarrow})$ as follows:
\vspace{-5pt}
\begin{equation}\small
\begin{aligned}
\vspace{-2pt}
\label{eq:tho_1}
\frac{\#(u_i,u_j)}{|\mathcal{D}|}\stackrel{p}{\rightarrow}\frac{1}{2\cdot ws}\sum_{r=1}^{ws}\left(\frac{d_{i}}{vol(G)}(\textbf{P}^r)_{ij}+\frac{d_{j}}{vol(G)}(\textbf{P}^r)_{ij}\right).\\
\end{aligned}
\end{equation}
\vspace{-2pt}
\begin{equation}\small
\begin{aligned}
\vspace{-2pt}
\label{eq:tho_1}
\frac{\#(u_i)}{|\mathcal{D}|}\stackrel{p}{\rightarrow}\frac{d_{i}}{vol(G)}.
\end{aligned}
\end{equation}
where $ws$ is the window size, $u_i$ is the center vertex and $u_j$ is a context vertex of $u_i$, $\#(u_i)$ and $\#(u_j)$ denote the frequencies of center vertex $u_i$ and context $u_j$ appeared in the corpus, respectively.
\end{lemma}
Lemma~\ref{lemma:l1}~\cite{DBLP:conf/wsdm/QiuDMLWT18} indicates that probabilities $\frac{\#(u_i,u_j)}{|\mathcal{D}|}$ and $\frac{\#(u_i)}{|\mathcal{D}|}$ can be computed based on the transition matrix of a random walk directly, rather than generating the vertex sequences.
Based on Lemma~\ref{lemma:l1}, we can further simplify to compute the proximity of center-context pair $(u_i, u_j)$ in corpus $\mathcal{D}$ as the following theorem:
\begin{theorem}
\label{theorem:1}
Let $\textbf{P}$ be the transition matrix of a random walk, and $\mathcal{D}$ be the generated corpus. When $L_\mathcal{D} \rightarrow \infty$, we have
\begin{equation}
\begin{aligned}
\frac{\#(u_i, u_j)}{\#(u_i)}&\stackrel{p} {\rightarrow} \frac{1}{ws}\sum_{r=1}^{ws}(\textbf{P}^r)_{ij}.
\end{aligned}
\end{equation}

\begin{proof}
Since $f(x) = \frac{1}{x}$ is a continuous function, thus
\[\frac{|\mathcal{D}|}{\#(u_i)}\stackrel{p}{\rightarrow}\frac{vol(G)}{d_{i}}.\]
According to Lemma~\ref{lemma:l1}, we have
\begin{equation}\small
\begin{aligned}
\label{eq:tho_2}
\frac{\#(u_i, u_j)}{\#(u_i)}&\stackrel{p}{\rightarrow}\frac{1}{2\cdot ws}\sum_{r=1}^{ws}\frac{vol(G)}{d_{i}}\left(\frac{d_{i}}{vol(G)}(\textbf{P}^r)_{ij}+\frac{d_{j}}{vol(G)}(\textbf{P}^r)_{ij}\right)\\
&=\frac{1}{2\cdot ws}\left(\sum_{r=1}^{ws}(\textbf{P}^r)_{ij}+\frac{d_{j}}{d_{i}}\sum_{r=1}^{ws}(\textbf{P}^r)_{ij}\right).
\end{aligned}
\end{equation}
Moreover, the above equation can be rewritten in the matrix form as follows:
\begin{equation}\small
\begin{aligned}
\label{eq:tho_m}
&\sum_{r=1}^{ws}(\textbf{P}^r)+\sum_{r=1}^{ws}\textbf{D}^{-1}(\textbf{P}^r)^T\textbf{D}\\
&=\sum_{r=1}^{ws}(\textbf{P}^r)+\sum_{r=1}^{ws}(\textbf{P}^r)=2\sum_{r=1}^{ws}(\textbf{P}^r).
\end{aligned}
\end{equation}
Thus, we can obtain $\frac{\#(u_i, u_j)}{\#(u_i)}\stackrel{p} {\rightarrow} \frac{1}{ws}\sum_{r=1}^{ws}(\textbf{P}^r)_{ij}$.
\end{proof}
\end{theorem}

Finally, each entry of matrix $\textbf{M}^U$ is approximated as:
\begin{eqnarray}\small
\textbf{M}^U_{ij} = (x_{ij}^U)^* = {\textbf{u}_i}^{T}{\pmb{\theta}_{j}} \approx \log{\frac{1}{ws \cdot q(u_i,u_j)}\sum_{r=1}^{ws}(\textbf{P}^U)^r_{ij}}-\log{ns}
.\label{eq:m_u} \nonumber
\end{eqnarray}

Similarly, each entry of matrix $\textbf{M}^V$ can be also approximated as:
\begin{eqnarray}\small
\textbf{M}^V_{ij} = (x_{ij}^V)^* = {\textbf{v}_i}^{T}{\pmb{\theta}_{j}} \approx \log{\frac{1}{ws \cdot q(v_i,v_j)}\sum_{r=1}^{ws}(\textbf{P}^V)^r_{ij}}-\log{ns}
.\label{eq:m_v} \nonumber
\end{eqnarray}

However, the matrices $\textbf{M}^U$ and $\textbf{M}^V$ are not only ill-defined since $\log{0}=-\infty$, but also they are dense, which leads to a computational challenge of factoring the matrices $\textbf{M}^U$ and $\textbf{M}^V$ by element-wise algorithms.
Inspiring by the shifted PPMI approach~\cite{DBLP:conf/nips/LevyG14}, we define $\textbf{M}_{ij}'^R=\max{(\textbf{M}_{ij}^R,0)}$, where $R\in\{U,V\}$.
In this way, $\textbf{M}'^U$ and $\textbf{M}'^V$ become sparse matrices.

\subsection{Co-factorizing Multiple Matrices}
To learn the representation of vertices $u_i$ and $v_i$, we only need to co-factorize the matrices $W$, $\textbf{M'}^U$ and $\textbf{M'}^V$.
%
%
As such, BiNE with LSH-based negative sampling can be transferred into matrix co-factorization method, namely BiNE-MF, which learns the vertex representation via employing the matrix co-factorization.

\begin{figure}[t]
	\centering
	\includegraphics[width=0.4\textwidth]{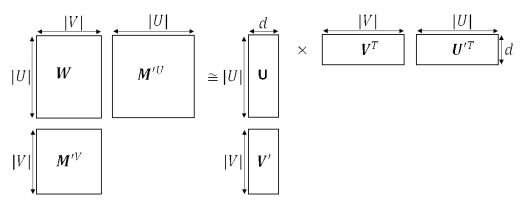}
	\vspace{-5pt}
	\caption{An example of implicitly factorizing multiple matrices}
	\label{fig:matrix factorization}
    \vspace{-15pt}
\end{figure}

In detail, BiNE-MF can be regarded as jointly factorizing three matrices: the weighted matrix $\textbf{W}$ and two proximity matrices $\textbf{M'}^U$, $\textbf{M'}^V$, where $\textbf{W}$ preserves the explicit relations in the bipartite network, and both $\textbf{M'}^U$ and $\textbf{M'}^V$ preserve the implicit relations in the bipartite network.
As shown in Fig.~\ref{fig:matrix factorization}, matrix $\textbf{M}'^U$ shares the vertex embedding matrix of $\textbf{U}$ with $\textbf{W}$, and matrix $\textbf{M}'^V$ shares the vertex embedding matrix of $\textbf{V}$ with $\textbf{W}$.
We set $\textbf{H}$ as 
\begin{eqnarray}
\textbf{H} = \left(
	\begin{matrix}
	\textbf{W} & \alpha'\textbf{M}'^U \\
    \beta'\textbf{M}'^V & \textbf{O}
	\end{matrix}
\right),\nonumber
\end{eqnarray}
where \textbf{O} can be any matrix (i.e., loss will be zero when matrix \textbf{O} is reconstructed), $\alpha'$ and $\beta'$ are scaling parameters to balance the importance of explicit and implicit relations. 
For BiNE-MF, we employ the KL-divergence to measure the difference between two distributions $p(i,j)$ and $\hat{p}(i,j)$, where they are
\begin{eqnarray}
P(i,j) = \frac{H_{ij}}{\sum_{s,t}H_{st}},
\hat{p}(i,j) = \frac{1}{1+exp(-{\textbf{u}_i}^T{\textbf{u}_j})}.
\end{eqnarray}

Once we obtain the matrix $\textbf{H}$, we can also employ symmetric SVD~\cite{DBLP:conf/nips/LevyG14} or SMF (stochastic matrix factorization)~\cite{DBLP:journals/computer/KorenBV09} to map the vertices into a low-dimensional space and obtain the embedding matrices $\textbf{U}$ and $\textbf{V}$.

\subsection{Discussions}
\label{subsec:LSH and SGNS}
\subsubsection{Effect of Negative Sampling}
As mentioned in Equation~(\ref{eq:conditional}), LSH-based negative sampling method used in BiNE is implicitly factorizing
\vspace{-5pt}
\begin{equation}
\label{eq:BiNE}
\log{\frac{\#(u_i,u_j)}{\#(u_i)\cdot q(u_i,u_j)}}-\log{ns}.
\end{equation}
where $q(u_i,u_j)$ is the probability that $u_i$ and $u_j$ locate in the different buckets of LSH.
In contrast, the frequency-based negative sampling method in  SGNS~\cite{DBLP:conf/nips/LevyG14} is implicitly factorizing
\vspace{-5pt}
\begin{equation}
\label{eq:SGNS}
\log{\frac{\#(u_i,u_j)\cdot|\mathcal{D}|}{\#(u_i)\cdot\#(u_j)}}-\log{ns}.
\end{equation}
Here, 
$\frac{\#(u_i,u_j)}{\#(u_i)} = p(u_j|u_i)$ represents the proximity of center-context pair $(u_i,u_j)$ in the corpus $\mathcal{D}$.
Moreover, both $q(u_i,u_j)$ and $\frac{\#(u_j)}{|\mathcal{D}|}$ can be treated as penalty factors which contribute to measure the proximity of center vertex with its context vertices.

Assume that $u_i$ and $u_j$ have higher proximity, i.e., $P(u_j|u_i)$ is larger. 
Then, $q(u_i,u_j)$ used in BiNE will be smaller due to the lower probability of mapping $u_i$ and $u_j$ into different buckets. 
In this case, $\log{\frac{\#(u_i,u_j)}{\#(u_i)\cdot q(u_i,u_j)}}-\log{ns}$ will be larger. However, $\frac{\#(u_j)}{|\mathcal{D}|}$ used in SGNS is undetermined since its value can be large or small. 
Furthermore, the large value of $\log{\frac{\#(u_i,u_j)\cdot|\mathcal{D}|}{\#(u_i)\cdot\#(u_j)}}-\log{ns}$ biases $u_j$ towards infrequent vertices, which may be inconsistent with our intuition.

\subsubsection{Computational Complexity Analysis}
Computing the matrices $\textbf{M}'^U$ (or $\textbf{M}'^V$) directly is a challenging task when the homogeneous network is large and dense or the window size is large.
We resort to the approximation algorithm proposed in the work~\cite{DBLP:conf/wsdm/QiuDMLWT18} to reduce the computation complexity.
Unlike the online training method BiNE, the MF-based methods, such as SMF and BiNE-MF, work over aggregated center-context pair statistics using $(u_i,v_j，w_{ij})$, $(u_i,u_j,\textbf{M}_{ij}'^U)$ and $(v_i,v_j,\textbf{M}_{ij}'^V)$ triples as input, which makes the optimization more directly and salable to large bipartite networks.

\vspace{-6pt}
\section{Experiments}
\label{sec:experiments}
In this section, we perform experiments on real-world datasets with the aim of answering the following research questions:
\begin{itemize}
	\item[\textbf{RQ1}] How does BiNE perform compared with state-of-the-art network embedding methods? 
    \item[\textbf{RQ2}]Can the MF-based methods, such as symmetric SVD, SMF, and BiNE-MF, achieve similar performance as that of BiNE? 
	\item[\textbf{RQ3}] Can our proposed random walk generator contribute to preserving the long-tail distribution of vertex in bipartite networks and the implicit relations mined by it be helpful to learn better vertex representations?
    \item[\textbf{RQ4}] Does LSH-based negative sampling strategies superior to the frequency-based strategies in modeling bipartite networks?
\end{itemize}
The following sections will illustrate the experimental settings before answering the above research questions.
In addition, a case study that visualizes a small bipartite network is performed to demonstrate the rationality of BiNE.
\subsection{Experimental Settings}
\subsubsection{\textbf{Datasets}}
We purposefully chosen unweighted networks for link prediction, which is usually approached as a classification task that predicts whether a link exists between two vertices;
while we use weighted networks for recommendation task, which is a personalized ranking task that aims to provide items of interest for a user.

\begin{enumerate}[leftmargin=*]
\item Unweighted bipartite network.
We construct two unweighted bipartite networks from Wikipedia and Tencent, respectively. Specifically, the Wikipedia dataset contains the edit relationship between authors and pages, which is public accessible\footnote{http://konect.uni-koblenz.de/networks/edit-enwiki}; The Tencent dataset records the watching behaviors of users on movies in QQlive\footnote{https://v.qq.com/} in one month's time.

\item Weighted bipartite network.
We construct other three weighted bipartite network from DBLP, Movielens and VisualizeUs, respectively. Specifically, the DBLP\footnote{http://dblp.uni-trier.de/xml/}  dataset contains the publish relationship between authors and venues, where the edge weight indicates the number of papers published on a venue by an author; The Movielens\footnote{http://grouplens.org/datasets/movielens/} dataset records the rating behavior of users on movies, where the edge weight denotes the rating score of a user on a movie; The VisualizeUs\footnote{http://konect.uni-koblenz.de/networks/pics\_ti} dataset records the tagging behavior of users on pictures, where the edge describes the number of times of tagging of a user on a picture.
\end{enumerate}
The statistics of our experimented datasets are summarized in Table~\ref{tab:statistics and metrics}.

\begin{table}[htbp]
\centering
\caption{Statistics of bipartite networks and metrics adopted in experiments for different tasks.}
\label{tab:statistics and metrics}
\resizebox{!}{12.5mm}{
\begin{tabular}{|l|r|r|r|r|r|}
\hline
Task    & \multicolumn{2}{c|}{Link Prediction}& \multicolumn{3}{c|}{Recommendation}  \\ \hline
Type   & \multicolumn{2}{c|}{undirected, unweighted} & \multicolumn{3}{c|}{undirected, weighted} \\ \hline
Metric   & \multicolumn{2}{c|}{AUC-ROC,AUC-PR} & \multicolumn{3}{c|}{F1, NDCG, MAP, MRR } \\ \hline
Name    & Tencent    & Wikipedia & VisualizeUs & DBLP & MovieLens      \\ \hline
$|U|$     & 14,259     & 15,000  & 6,000     & 6,001    & 69,878            \\ \hline
$|V|$     & 1,149      & 3,214   & 3,355     & 1,308    & 10,677              \\
\hline
$|E|$     & 196,290    & 172,426 & 35,639    &  29,256      & 10,000,054     \\
\hline
Density   & 1.2\%     & 0.4\% & 0.2\%      &  0.4\%      & 1.3\%            \\ \hline
\end{tabular}}
\vspace{-5pt}
\end{table}
\vspace{-5pt}
\subsubsection{\textbf{Evaluation Protocols}}
\begin{enumerate}[leftmargin=*]
\item For link prediction task, we first apply the same protocol as the Node2vec paper~\cite{DBLP:conf/kdd/GroverL16} to process the Wikipedia dataset. Specifically, we first treat the observed links as the positive instances, and sample an equal number of unconnected vertex pairs as the negative instances. 
For Tencent dataset, we treat the user-movie pairs as positive instances if the user has watched the movie for more than 5 minutes, otherwise, negative instances.
For both datasets, we randomly sample $60\%$ instances as the training set, and evaluate the link prediction performance on the remaining $40\%$ dataset with two metrics: the ROC curve (AUC-ROC) and Precision-Recall curve (AUC-PR).
\item  For recommendation task, we randomly sample $60\%$ edges as the training data, and the remaining $40\%$ edges are treated as testing dataset for all datasets.
We rank all items in the testing set for each user and truncate the ranking list at 10 to evaluate the performance of top-10 recommendation with four IR metrics: F1, Normalized Discounted Cumulative Gain (NDCG), Mean Average Precision (MAP), and Mean Reciprocal Rank (MRR).
\end{enumerate}
For each metric, we compute the average score for all users, and perform paired sample T-test on it.
To avoid over-fitting, we generate 10 folds of train-test split, tuning hyper-parameters on the first fold only for each method.
We use the optimal hyper-parameter setting and report the average performance of all folds (i.e., the score of each metric and the p-value of t-test).
\vspace{-5pt}
\subsubsection{\textbf{Baselines}}
We compare BiNE with three types of baselines:
\begin{enumerate}[leftmargin=*]
\item \textbf{Network Embedding Methods}. 
We chose four representative of state-of-the-art network embedding methods as our baselines, including homogeneous and heterogeneous network embedding methods.
For each method, we use the released implementations for our experiments.
    \begin{itemize}[leftmargin=*]
    \item DeepWalk~\cite{DBLP:conf/kdd/PerozziAS14}: This method performs uniform random walks to get a corpus of vertex sequences. Then the word2vec is applied on the corpus to learn vertex embeddings for homogeneous networks.
    \item Node2vec~\cite{DBLP:conf/kdd/GroverL16}: This approach extends DeepWalk by performing biased random walks to generate the corpus of vertex sequences. The hyper-parameters $p$ and $q$ are set to 0.5 which has empirically shown good results.
    \item LINE~\cite{DBLP:conf/www/TangQWZYM15}: In contrast to  the above methods, this approach optimizes both the 1st-order and 2nd-order proximities in a homogeneous network without generating corpus. We use the LINE (1st+2nd) method which has shown the best performance in their paper.
    \item Metapath2vec++~\cite{dong2017metapath2vec}: As a state-of-the-art method for embedding heterogeneous networks, it generates corpus following the predefined meta-path scheme.
    And the meta-path scheme chosen in our experiments are ``IUI'' (item-user-item) and ``IUI''+``UIU'' (user-item-user), and we only report the best result between them.
    \end{itemize}

\item We compare with a set of methods that are specifically designed for the link prediction task. 
We apply several indices proposed in \cite{DBLP:conf/asunam/XiaDLZX12}, including
Absent Links (AL),
Katz Index (Katz), and Preferential Attachment (PA).

\item We also compare with several competitive methods\footnote{We use the LibRec implementation: \url{https://www.librec.net/}} that are designed for the top-K item recommendation task.
    \begin{itemize}[leftmargin=*]
    \item BPR~\cite{DBLP:conf/uai/RendleFGS09}:This method has been widely used in recommendation literature as a highly competitive baseline~\cite{DBLP:conf/www/HeLZNHC17}. It optimizes the matrix factorization (MF) model with a pairwise ranking-aware objective. 
    \item RankALS~\cite{DBLP:conf/recsys/TakacsT12}: This method also optimizes the MF model for the ranking task, by towards a different pairwise regression-based loss.
    \item FISMauc~\cite{DBLP:conf/kdd/KabburNK13}: Distinct to MF, factored item similarity model (FISM) is an item-based collaborative filtering method. We employ the AUC-based objective to optimize FISM for the top-K task.
    \item IRGAN~\cite{DBLP:conf/sigir/WangYZGXWZZ17}: This method combines two types of information retrieval models into the adversarial training. The generative model generates negative items for each user, the discriminative model distinguishes positive and negative items as far as possible.
\end{itemize}
\end{enumerate}
\vspace{-10pt}
\begin{table}[htbp]
\small\centering
\caption{The search range and optimal setting (highlighted in red) of hyper-parameters for our BiNE method.}
\vspace{-5pt}
\label{tab:parameter settings}
\resizebox{!}{12.9mm}{
\begin{tabular}{|c|l|l|}
\hline
\textbf{Parameter} & \multicolumn{1}{c|}{\textbf{Meaning}}      & \multicolumn{1}{c|}{\textbf{Test values}}          \\ \hline\hline
$ns$      & number of negative samples        & {[}1, 2, {\color{red}{\textbf{4}}}, 6, 8, 10{]}                   \\ \hline
$ws$      & size of window                    & {[}1, 3,  {\color{red}{\textbf{5}}}, 7, 9{]}                       \\ \hline
$p$       & walk stopping probability             & {[}0.05, 0.1,  {\color{red}{\textbf{0.15}}}, 0.2, 0.3, 0.4, 0.5{]} \\ \hline
$\beta$   &  trade-off  parameter      & {[}0.0001, 0.001,  {\color{red}{\textbf{0.01}}}, 0.1, 1{]}         \\ \hline
$\gamma$   & trade-off  parameter      & {[}0.01, 0.05,  {\color{red}{\textbf{0.1}}}, 0.5,  {\color{red}{\textbf{1}}}, 5{]}          \\ \hline\hline
$\alpha'$   &  trade-off  parameter      & {[}0.0001, {\color{red}{\textbf{0.001}}},  {\color{red}{\textbf{0.01}}},  {\color{red}{\textbf{0.1}}}, 1{]}         \\ \hline
$\beta'$   & trade-off  parameter      & {[}0.0001, {\color{red}{\textbf{0.001}}},  {\color{red}{\textbf{0.01}}},  {\color{red}{\textbf{0.1}}}, 1{]}            \\ \hline
\end{tabular}}
\vspace{-5pt}
\end{table}

\subsubsection{\textbf{Parameter Settings}}
We have fairly tuned the hyper-parameters for each method.
For all network embedding methods, we set the embedding size as 128 for a fair comparison; other hyper-parameters follow the default setting of their released implementations.
For the recommendation baselines, we tuned the learning rate and latent factor number since they impact most on the performance; other hyper-parameters follow the default setting of the LibRec toolkit.

For BiNE, we fix the loss trade-off parameter $\alpha$ as 0.01 and tune the other two.
The $minT$ and $maxT$ are respectively set to 1 and 32, which empirically show good results in most cases.
We test the learning rate $\lambda$ of [0.01, 0.025, 0.1].
And the optimal setting of learning rate is $0.025$ for the VisualizeUs/DBLP dataset and $0.01$ for others.
The search range and optimal setting (highlighted in red font) of other parameters are shown in Table~\ref{tab:parameter settings}.
Note that besides $\gamma$ is set differently --- 0.1 for recommendation and 1 for link prediction --- other parameters are set to the same value for both tasks.

For different MF-based methods, we set $\alpha'$ and $\beta'$ as the same values for pair comparison.
And the optimal setting of $\alpha'$ and $\beta'$ are $0.001$ and $0.001$ for VisualizeUs/DBLP/Wikipedia dataset, $0.01$ and $0.01$ for Tencent dataset and $0.1$ and $0.1$ for Movielens.
We test the learning rate of [0.00001, 0.00005, 0.0001, 0.0005, 0.001, 0.005, 0.01] for SMF and BiNE-MF.
\subsection{Performance Comparison (RQ1)}
\begin{center}
\setlength{\tabcolsep}{1.6mm}{
\begin{table}[!htpb]
\centering{
\caption{Link prediction performance on Tencent and Wikipedia.
}
\vspace{-5pt}
\label{tab:link prediction}
\begin{tabular}{|l|r|r||r|r|}
\hline
\multirow{2}{*}{\textbf{Algorithm}} & \multicolumn{2}{c||}{\textbf{Tencent}}                          & \multicolumn{2}{c|}{\textbf{Wikipedia}}                        \\ \cline{2-5}
                           & \multicolumn{1}{c|}{\textbf{AUC-ROC}} & \multicolumn{1}{c||}{\textbf{AUC-PR}} & \multicolumn{1}{c|}{\textbf{AUC-ROC}} & \multicolumn{1}{c|}{\textbf{AUC-PR}} \\ \hline\hline
\textbf{AL}           & 50.44\%                       & 65.70\%                      & 90.28\%                       & 91.81\%                      \\
\textbf{Katz}         & 50.90\%                       & 65.06\%                      & 90.84\%                       & 92.42\%                      \\
\textbf{PA}                         & 55.60\%                       & 68.99\%                      & 90.71\%                       & 93.37\%                      \\
\hline\hline
\textbf{DeepWalk}                   & 57.62\%                       & 71.32\%                      & 89.71\%                       & 91.20\%                      \\
\textbf{LINE}                       & 59.68\%                       & 73.48\%                      & 91.62\%                       & 93.28\%                      \\
\textbf{Node2vec}                   & 59.28\%                       & 72.62\%                      & 89.93\%                       & 91.23\%                      \\
\textbf{Metapath2vec++}             & 60.70\%                       & 73.69\%                      & 89.56\%                       & 91.72\%                      \\\hline\hline
\textbf{BiNE}                       & \textbf{60.98\%**}              & \textbf{73.77\%**}             & \textbf{92.91\%**}              & \textbf{94.45\%**}            \\\hline
\end{tabular}}
\begin{tablenotes}
\centering
\small
\item[1] ** indicates that the improvements are statistically significant for $p<0.01$ judged by paired t-test.
\end{tablenotes}
\vspace{-10pt}
\end{table}}
\end{center}
\begin{center}
\setlength{\tabcolsep}{0.6mm}{
\begin{table*}[!htpb]
\centering{
\caption{Performance comparison of Top-10 recommendation on VisualizeUs, DBLP, and MovieLens.
}
\vspace{-5pt}
\label{tab:movie}
\begin{tabular}{|l|r|r|r|r||r|r|r|r||r|r|r|r|}
\hline
\multirow{2}{*}{\textbf{Algorithm}} & \multicolumn{4}{c||}{\textbf{VisualizeUs}}                                                                                                     & \multicolumn{4}{c||}{\textbf{DBLP}}                                                                                                & \multicolumn{4}{c|}{\textbf{Movielens}}                                                                                                  \\ \cline{2-13}
                                    & \multicolumn{1}{c|}{\textbf{F1@10}} & \multicolumn{1}{c|}{\textbf{NDCG@10}} & \multicolumn{1}{c|}{\textbf{MAP@10}} & \multicolumn{1}{c||}{\textbf{MRR@10}} & \multicolumn{1}{c|}{\textbf{F1@10}} & \multicolumn{1}{c|}{\textbf{NDCG@10}} & \multicolumn{1}{c|}{\textbf{MAP@10}} & \multicolumn{1}{c||}{\textbf{MRR@10}} & \multicolumn{1}{c|}{\textbf{F1@10}} & \multicolumn{1}{c|}{\textbf{NDCG@10}} & \multicolumn{1}{c|}{\textbf{MAP@10}} & \multicolumn{1}{c|}{\textbf{MRR@10}} \\ \hline\hline
\textbf{BPR}                        & 6.22\%                           & 9.52\%                             & 5.51\%                            & 13.71\%                           & 8.95\%                           & 18.38\%                            & 13.55\%                           & 22.25\%                           & 8.03\%                           & 7.58\%                             & 2.23\%                            & 40.81\%                           \\
\textbf{RankALS}                    & 2.72\%                           & 3.29\%                             & 1.50\%                            & 3.81\%                            & 7.62\%                           & 11.50\%                            & 7.52\%                            & 14.87\%                           & 8.48\%                           & 7.95\%                             & 2.66\%                            & 38.93\%                           \\
\textbf{FISMauc}                    & 10.25\%                          & 15.46\%                            & 8.86\%                            & 16.67\%                           & 9.81\%                           & 13.77\%                            & 7.38\%                            & 14.51\%                           & 6.77\%                           & 6.13\%                             & 1.63\%                            & 34.04\%                           \\ 

\textbf{IRGAN}                    & 12.91\%                          & 22.16\%                            & 14.37\%                            & 29.02\%                           & 11.20\%                           & 22.60\%                            & 16.57\%                            & 28.17\%                           & 6.79\%                           & 6.91\%                             & 2.19\%                            & 41.19\%                     \\ \hline\hline 
\textbf{DeepWalk}                   & 5.82\%                           & 8.83\%                             & 4.28\%                            & 12.12\%                           & 8.50\%                           & 24.14\%                            & 19.71\%                           & 31.53\%                           & 3.73\%                           & 3.21\%                             & 0.90\%                            & 15.40\%                           \\
\textbf{LINE}                       & 9.62\%                           & 13.76\%                            & 7.81\%                            & 14.99\%                           & 8.99\%                           & 14.41\%                            & 9.62\%                            & 17.13\%                           & 6.91\%                           & 6.50\%                             & 1.74\%                            & 38.12\%                           \\
\textbf{Node2vec}                   & 6.73\%                           & 9.71\%                             & 6.25\%                            & 13.95\%                           & 8.54\%                           & 23.89\%                            & 19.44\%                           & 31.11\%                           & 4.16\%                           & 3.68\%                             & 1.05\%                            & 18.33\%                           \\
\textbf{Metapath2vec++}             & 5.92\%                           & 8.96\%                             & 5.35\%                            & 13.54\%                           & 8.65\%                           & 25.14\%                            & 19.06\%                           & 31.97\%                           & 4.65\%                           & 4.39\%                             & 1.91\%                            & 16.60\%                           \\ \hline\hline
\textbf{BiNE}                       & \textbf{13.63\%**}                 & \textbf{24.50\%**}                   & \textbf{16.46\%**}                  & \textbf{34.23\%**}                  & \textbf{11.37\%**}                 & \textbf{26.19\%**}                   & \textbf{20.47\%**}                  & \textbf{33.36\%**}                  & \textbf{9.14\%**}                  & \textbf{9.02\%**}                  & \textbf{3.01\%**}                   & \textbf{45.95\%**}                  \\\hline
\end{tabular}}

\begin{tablenotes}
\centering
\small
\item[1] ** indicates that the improvements are statistically significant for $p<0.01$ judged by paired t-test.
\end{tablenotes}
\end{table*}
}
\end{center}
\subsubsection{\textbf{Link Prediction}}
In this task,  we first concatenate the embedding vectors ${\textbf{u}_i}$, ${\textbf{v}_j}$ and label (\ie the label of the positive instance is 1, otherwise 0) as a record for each instance $(u_i, v_j)$ in the dataset, then feed the record into the logistic regression classifier with L2 loss function.
Table~\ref{tab:link prediction} illustrates the performance of baselines and our BiNE, where we have the following key observations:
\begin{itemize}[leftmargin=*]
\item The neural network-based methods which trained in a data-dependent supervised manner outperform the indices proposed in~\cite{DBLP:conf/asunam/XiaDLZX12} significantly.
\item Metapath2vec++ and BiNE are significantly better than other neural network-based methods. This improvement demonstrates the positive effect of considering the information of node types when embedding bipartite networks.
\item BiNE outperforms Metapath2vec++ significantly and achieves the best performance on both datasets in both metrics. 
This improvement points out the effectiveness of modeling of explicit and implicit relations in different ways.
\end{itemize}
%
\subsubsection{\textbf{Recommendation}}
In this task, we adopt the inner product kernel ${\textbf{u}_i}^T{\textbf{v}_j}$ to estimate the preference of user $u_i$ on item $v_j$, and evaluate performance on the top-10 results.
Table~\ref{tab:movie} shows the performance of baselines and our BiNE, where we have the following key observations:
\begin{itemize}[leftmargin=*]
\item BiNE outperforms all baselines on all datasets, and the improvements are significant compared with  Metapath2vec++ though it  also considers the node type information when embedding bipartite networks. We hold that it is due to it ignores the weights and treats the two types of relations (\ie the explicit and implicit relations) as equally.
\item BiNE outperforms LINE significantly though it also consider the weight information when embedding networks. The suboptimal performance obtained by LINE because of two reasons.
(1) LINE ignores further high-order proximities among vertices due to it only preserves both 1st-order and 2nd-order relations when learning the representations for vertices.
(2) LINE learns two separated embeddings for 1st-order and 2nd-order relations and concatenates them via post-processing, rather than optimizing them in a unified framework.
Whereas BiNE mines high-order implicit relations among homogeneous vertices by performing random walks and design a joint framework to model the explicit and implicit relations jointly, where different relations reinforce each other and lead to better vertex representations.

\item Although IRGAN performs the best among all the baselines, BiNE outperforms IRGAN significantly. 
This is due to the factor that IRGAN takes the user-item interaction as positive samples and generates negative samples by adversarial training, which not fully captures the global structure of the network.
This points to the advantage of modeling both explicit and implicit relations into the embedding process.

\end{itemize}

\subsection{Performance of MF-based Methods (RQ2)}
\label{subsec:RQ2}
\begin{center}
\setlength{\tabcolsep}{3.5mm}{
\begin{table*}[t]
\centering{
\caption{Performance comparison of matrix factorization using different manners. }
\vspace{-5pt}
\label{tab:MF}
\begin{tabular}{|l|r|r||r|r||r|r||r|r|}
\hline
                     & \multicolumn{2}{c||}{\textbf{Symmetric SVD}}  & \multicolumn{2}{c||}{\textbf{SMF}}  & \multicolumn{2}{c||}{\textbf{BiNE-MF}} & \multicolumn{2}{c|}{\textbf{BiNE}}\\ \hline\hline
                     & \multicolumn{8}{c|}{\textbf{Link Prediction}}                                                                \\ \hline\hline
\textbf{Dataset}     & \textbf{AUC-ROC} & \textbf{AUC-PR} & \textbf{AUC-ROC} & \textbf{AUC-PR} & \textbf{AUC-ROC} & \textbf{AUC-PR} & \textbf{AUC-ROC} & \textbf{AUC-PR}\\ \hline
\textbf{Tencent}     &  62.58\%                &75.08\%                 &  \textbf{62.81\%}                &  \textbf{75.19\%}               & 61.39\%          & 74.12\%    &60.98\%	& 73.77\%   \\ 
\textbf{WikiPedia}   & 87.55\%                 & 90.93\%                &  91.27\%                &  93.68\%               & 92.17\%          & \textbf{94.51\%}    &\textbf{92.91\%**}	&  94.45\%   \\ \hline\hline
\textbf{}            & \multicolumn{8}{c|}{\textbf{Recommendation}}                                                                 \\ \hline
\textbf{Dataset}     & \textbf{MAP@10}  & \textbf{MRR@10} & \textbf{MAP@10}  & \textbf{MRR@10} & \textbf{MAP@10}  & \textbf{MRR@10} & \textbf{MAP@10}  & \textbf{MRR@10}\\ \hline
\textbf{VisualizeUs} & 2.47\%                 & 4.70\%                & 14.26\%                 & 32.04\%                & 16.27\%          & 34.02\%    &\textbf{16.46\%**}	&\textbf{34.23\%**}     \\ 
\textbf{DBLP}        & 2.44\%                 & 5.33\%                & 10.19\%                 & 21.54\%                & 19.99\%          & \textbf{33.40\%}     &\textbf{20.47\%**}	& 33.36\%   \\ 
\textbf{Movielens}   & 0.33\%                 &  3.49\%               &  2.25\%                & 41.91\%                & \textbf{3.26\%**}           & 45.17\%    &3.01\%	& \textbf{45.95\%**}   \\ \hline
\end{tabular}}

\begin{tablenotes}
\centering
\small
\item[1] ** indicates that the improvements are statistically significant for $p<0.01$ judged by paired t-test.
\end{tablenotes}
\end{table*}
}
\end{center}

Here we adopt symmetric SVD, stochastic matrix factorization (SMF) and BiNE-MF to obtain vertex embeddings and compare their performance.
The result is shown in Table~\ref{tab:MF}. We have the following key observations:
\begin{itemize}[leftmargin=*]
\item SMF and BiNE-MF that much like BiNE's training process show roughly equivalent performance with BiNE
.
They also obtain better performance than BiNE in the three big datasets: Tencent, WikiPedia, and Movielens.
This reveals one advantage of MF-based methods that they approximate the global implicit relations while the limited scale of random walk negatively impacts BiNE's performance.
And this observation is in line with the work\cite{DBLP:conf/wsdm/QiuDMLWT18}.
It also indicates that $maxT$ should be specified to a large number for a large-scale network.
\item Symmetric SVD yields the worser result than SMF and BiNE-MF in most case. The cause of this result is the matrix $\textbf{H}$ is sparse and symmetric SVD is undefined when the matrix is incomplete. In addition, addressing the relatively few observed entries is highly prone to overfitting~\cite{DBLP:journals/computer/KorenBV09}.
SMF and BiNE-MF are better than symmetric SVD at handling missing entries. And the regularization in SMF~\cite{DBLP:conf/nips/LevyG14} and nonlinear variation in BiNE-MF are also two workable ways of improving performance. 
\end{itemize}

\subsection{Utility of Random Walk Generator (RQ3)}
In this section, we first demonstrate the effectiveness of our random walk generator on preserving the characteristics of bipartite networks, especially the power-law distribution of vertices. Then, we illustrate the effect of considering the implicit relations when learning the representations of vertices for bipartite networks.
\begin{figure}[htbp]
	\centering
	\begin{subfigure}[htpb]{0.22\textwidth}
		\centering
		\includegraphics[width=\textwidth]{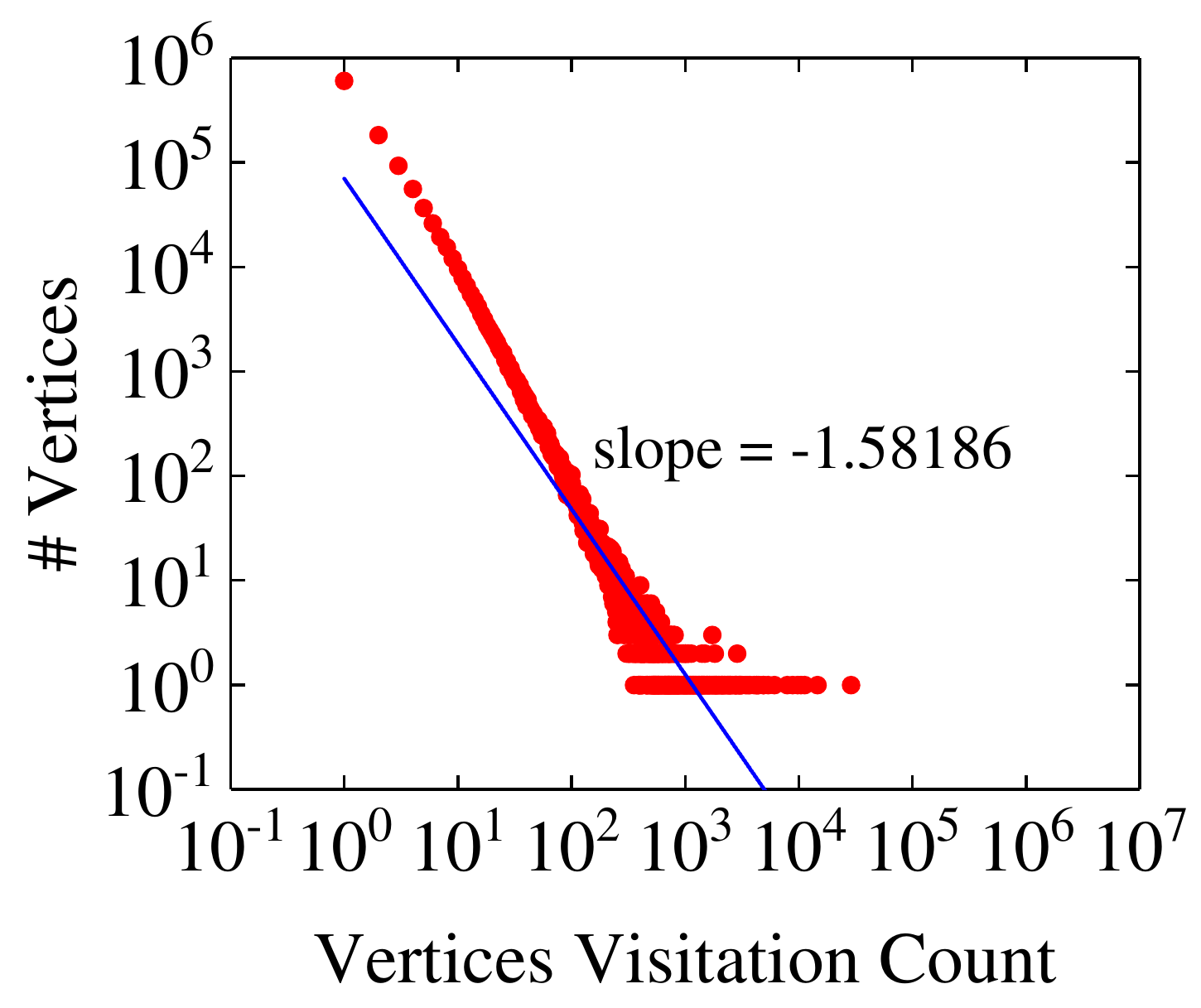}
		\vspace{-15pt}
		\caption{YouTube}
		\label{fig:deep_difficulty}
	\end{subfigure} 
	\begin{subfigure}[htpb]{0.22\textwidth}
		\centering
		\includegraphics[width=\textwidth]{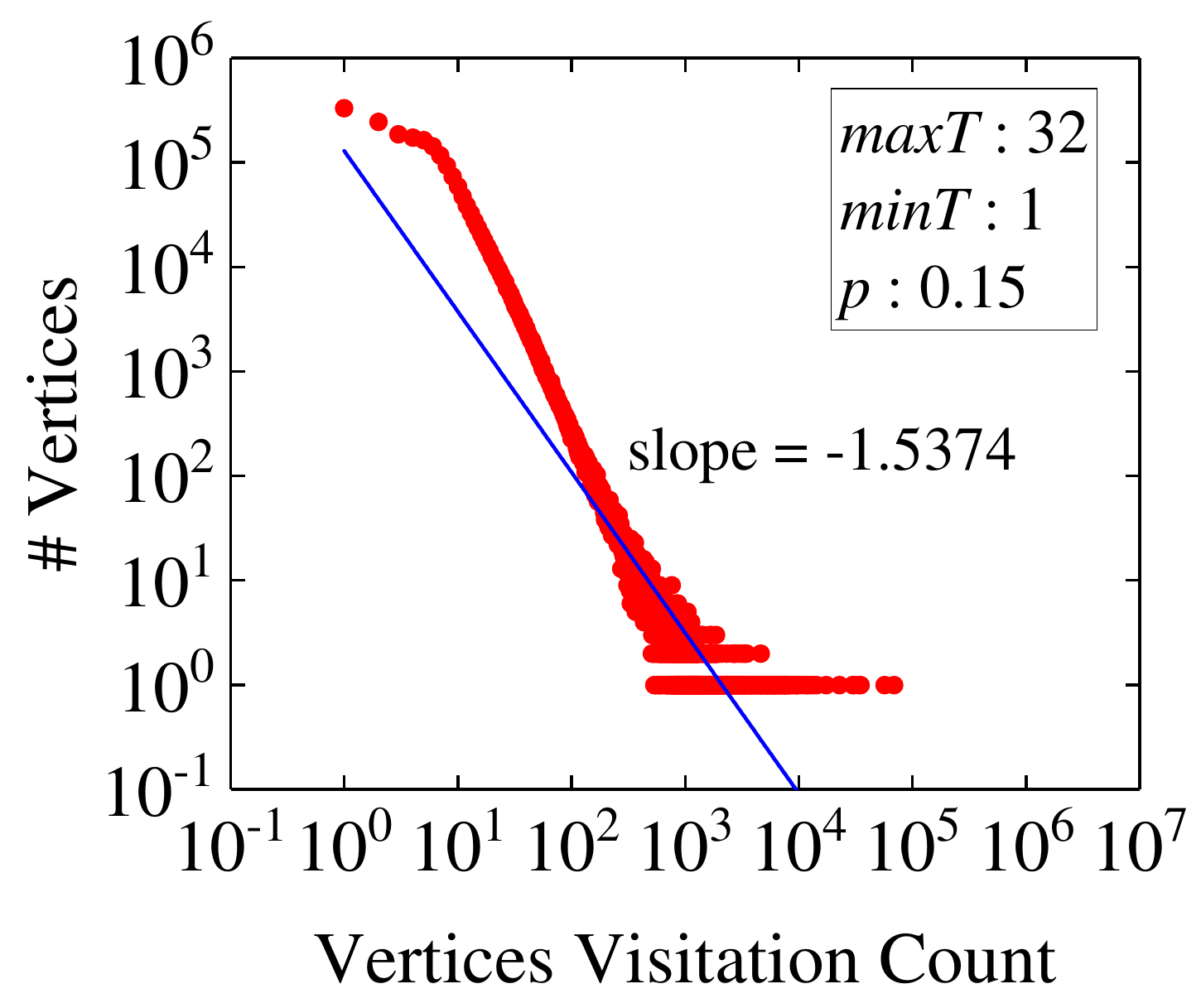}
		\vspace{-15pt}
		\caption{Random walk generator}
		\label{fig:deep_difficulty_pretrain}
	\end{subfigure} 
	\vspace{-5pt}
	\caption{The vertex distribution of (a) the real-world YouTube dataset and (b) the corpus generated by our biased and self-adaptive random walk generator.}
	\label{fig:walk}
\end{figure}

The frequency distribution of vertices in a real YouTube dataset is plotted in Fig.~\ref{fig:walk}(a). We can see that the vertices exhibit a standard power-law distribution with a slope of $-1.582$. By contrast, we plot the frequency distribution of vertices in a corpus obtained from our random walk generator in Fig.~\ref{fig:walk}(b). 
We can easily find that our random walk generator almost generates a standard power-law distribution with a slope $-1.537$ which is very close to that of the original network.

\begin{table}[htpb]
\small\centering
\caption{BiNE with different random walk generators.}
\vspace{-5pt}
\label{tab: random walks generator}
\resizebox{!}{19mm}{
\begin{tabular}{|l|r|r||r|r|}
\hline
                 & \multicolumn{2}{c||}{\begin{tabular}[c]{@{}c@{}}\textbf{Uniform Random} \\ \textbf{Walk Generator}\end{tabular}} & \multicolumn{2}{c|}{\begin{tabular}[c]{@{}c@{}}\textbf{Biased and Self-adaptive}\\ \textbf{Random Walk Generator}\end{tabular}} \\ \hline\hline
\multicolumn{5}{|c|}{\textbf{Link Prediction}}                                                                                                                                                                               \\ \hline
\textbf{Dataset}          & \multicolumn{1}{c|}{\textbf{AUC-ROC}}                   & \multicolumn{1}{c||}{\textbf{AUC-PR}}                    & \multicolumn{1}{c|}{\textbf{AUC-ROC}}                   & \multicolumn{1}{c|}{\textbf{AUC-PR}}                   \\ \hline
\textbf{Tencent}   & 59.75\%                                        & 73.06\%                                        & \textbf{60.98\%**}                               & \textbf{73.77\%**}                              \\
\textbf{WikiPedia} & 88.77\%                                        & 91.91\%                                        & \textbf{92.91\%**}                               & \textbf{94.45\%**}                              \\ \hline\hline

\multicolumn{5}{|c|}{\textbf{Recommendation}}                                                                                                                                                                                \\ \hline
\textbf{Dataset}          & \multicolumn{1}{c|}{\textbf{MAP@10}}                  & \multicolumn{1}{c||}{\textbf{MRR@10}}                  & \multicolumn{1}{c|}{\textbf{MAP@10}}                  & \multicolumn{1}{c|}{\textbf{MRR@10}}                 \\ \hline
\textbf{VisualizeUS}  & 15.93\%                                         & 33.66\%                                         & \textbf{16.46\%**}                                & \textbf{34.23\%**}                               \\
\textbf{DBLP} & 11.79\%                                        & 23.41\%                                        & \textbf{20.47\%**}                               & \textbf{33.36\%**}                              \\
\textbf{MovieLens} & 2.91\%                                         & 46.12\%                                         & \textbf{3.04\%**}                                & \textbf{46.20\%**}                               \\ \hline
\end{tabular}}
\begin{tablenotes}
\centering
\small
\item[1]** indicates that the improvements are statistically significant for $p<0.01$ judged by paired t-test.
\end{tablenotes}
\end{table}
In addition, we compare the performance of BiNE under two settings --- use or not use our proposed random walk generator.
As shown in Table~\ref{tab: random walks generator}, the biggest absolute improvements of BiNE using our proposed random walk generator are 4.14\% and 10.25\% for link prediction and recommendation, respectively.
The above result indicates that the biased and self-adaptive random walk generator is helpful to capture the power-law distribution of vertices and contributes to improving the vertex representations for embedding bipartite networks.
Please note that we change the value of $maxT$ to 128 for this empirical study on Movielens dataset because of  the default value may be to small to fully preserve the implicit relations for such a large-scale bipartite network.

Lastly, we show the performance of BiNE and its variant which ignores the implicit relations.
Due to the space limitation, we only show the performance on the recommendation task from two metrics: MAP@10 and MRR@10.
From Table~\ref{tab: implicit relations}, we can find that the largest absolute improvements of BiNE with implicit relations are 1.44\% and 18.58\% for link prediction and recommendation, respectively.
It demonstrates that our proposed way of mining high-order implicit relation as the complement of explicit relations is effect to modeling bipartite networks.
\begin{table}[t]
\small\centering
\caption{BiNE with and without implicit relations.}
\vspace{-5pt}
\label{tab: implicit relations}
\resizebox{!}{19mm}{
\begin{tabular}{|l|r|r||r|r|}
\hline
                 & \multicolumn{2}{c||}{\begin{tabular}[c]{@{}c@{}}\textbf{Without Implicit}\\ \textbf{Relations}\end{tabular}} & \multicolumn{2}{c|}{\begin{tabular}[c]{@{}c@{}}\textbf{With Implicit}\\   \textbf{Relations}\end{tabular}} \\ \hline\hline
\multicolumn{5}{|c|}{\textbf{Link Prediction}}                                                                                                                                                                   \\ \hline
\textbf{Dataset}          & \multicolumn{1}{c|}{\textbf{AUC-ROC}}                & \multicolumn{1}{c||}{\textbf{AUC-PR}}                 & \multicolumn{1}{c|}{\textbf{AUC-ROC}}                & \multicolumn{1}{c|}{\textbf{AUC-PR}}                \\ \hline
\textbf{Tencent}   & 59.78\%                                     & 73.05\%                                     & \textbf{60.98\%**}                            & \textbf{73.77\%**}                           \\
\textbf{WikiPedia} & 91.47\%                                     & 93.73\%                                     & \textbf{92.91\%**}                            & \textbf{94.45\%**}                           \\ \hline\hline

\multicolumn{5}{|c|}{\textbf{Recommendation}}                                                                                                                                                                    \\ \hline
\textbf{Dataset}          & \multicolumn{1}{c|}{\textbf{MAP@10}}               & \multicolumn{1}{c||}{\textbf{MRR@10}}               & \multicolumn{1}{c|}{\textbf{MAP@10}}               & \multicolumn{1}{c|}{\textbf{MRR@10}}              \\ \hline
\textbf{VisualizeUS}  & 9.10\%                                      & 19.76\%                                      & \textbf{16.46\%**}                             & \textbf{34.23\%**}                            \\
\textbf{DBLP} & 20.20\%                                      & 32.95\%                                     & \textbf{20.47\%**}                            & \textbf{33.36\%**} \\
\textbf{MovieLens} & 2.86\%                                      & 43.98\%                                      & \textbf{3.01\%**}                             & \textbf{45.95\%**}                            \\ \hline
 	\end{tabular}}
\begin{tablenotes}
\centering
\small
\item[1] ** indicates that the improvements are statistically significant for $p<0.01$ judged by paired t-test.
\end{tablenotes}
\end{table}

\subsection{Negative Sampling Strategies (RQ4)}
\label{subsec:RQ4}
We have analyzed the difference between LSH-based and frequency-based negative sampling methods in section~\ref{subsec:LSH and SGNS}, that is LSH-based method resorts to dissimilar information deriving from observed links to obtain more accurate proximity of center-context vertex pairs, while frequency-based method utilizes frequency information to lower the proximity of center vertex with context vertices having high frequency.
Here, we compare the performance of BiNE with different negative sampling strategies.

As shown in Table~\ref{tab: negative sampling strategies}, there is a slight advantage in LSH-based by comparing it with frequency-based negative sampling method.
%
%
Thus, we hold that LSH-based sampling method, which uses dissimilar information obtained from user behavior data, can generate more reasonable negative samples in modeling user behavior.

Frequency-based method also shows roughly equivalent  performance in most cases.
An intuitive explanation is that the number of negatives is large and the probability of sampling similar vertices as negative is small via frequency-based method strategies.
%

\begin{table}[htbp]
\small\centering
\caption{BiNE with different negative sampling strategies.}
\label{tab: negative sampling strategies}
\resizebox{!}{19.5mm}{
\begin{tabular}{|l|r|r||r|r|}
\hline
                 & \multicolumn{2}{c||}{\begin{tabular}[c]{@{}c@{}}\textbf{Frequency-based}\\ \textbf{Negative Sampling}\end{tabular}} & \multicolumn{2}{c|}{\begin{tabular}[c]{@{}c@{}}\textbf{LSH-based}\\   \textbf{Negative Sampling}\end{tabular}} \\ \hline\hline
\multicolumn{5}{|c|}{\textbf{Link Prediction}}                                                                                                                                                                   \\ \hline
\textbf{Dataset}          & \multicolumn{1}{c|}{\textbf{AUC-ROC}}                & \multicolumn{1}{c||}{\textbf{AUC-PR}}                 & \multicolumn{1}{c|}{\textbf{AUC-ROC}}                & \multicolumn{1}{c|}{\textbf{AUC-PR}}                \\ \hline
\textbf{Tencent}   & 60.80\%                                     & 73.64\%                                     & \textbf{60.98\%}                            & \textbf{73.77\%}                           \\
\textbf{WikiPedia} & 92.21\%                                     & 94.12\%                                     & \textbf{92.91\%**}                            & \textbf{94.45\%**}                           \\ \hline\hline

\multicolumn{5}{|c|}{\textbf{Recommendation}}                                                                                                                                                                    \\ \hline
\textbf{Dataset}          & \multicolumn{1}{c|}{\textbf{MAP@10}}               & \multicolumn{1}{c||}{\textbf{MRR@10}}               & \multicolumn{1}{c|}{\textbf{MAP@10}}               & \multicolumn{1}{c|}{\textbf{MRR@10}}              \\ \hline
\textbf{VisualizeUS}  & 15.07\%                                      & 32.27\%                                      & \textbf{16.46\%**}                             & \textbf{34.23\%**}                            \\
\textbf{DBLP} & 20.46\%                                      & 32.93\%                                     & \textbf{20.47\%}                            & \textbf{33.36\%**} \\
\textbf{MovieLens} & \textbf{3.01\%}
                                      & 45.86\%                                      & \textbf{3.01\%}                             & \textbf{45.95\%}                            \\ \hline
 	\end{tabular}}
\begin{tablenotes}
\centering
\small
\item[1] ** indicates that the improvements are statistically significant for $p<0.01$ judged by paired t-test.
\end{tablenotes}
\end{table}

\subsection{Case Study}
In this section, we perform a visualization study for a small bipartite network to illustrate that the rationality of our BiNE method.
\begin{figure}[htbp]
	\centering
	\begin{subfigure}[b]{0.16\textwidth}
		\centering
		\includegraphics[width=\textwidth]{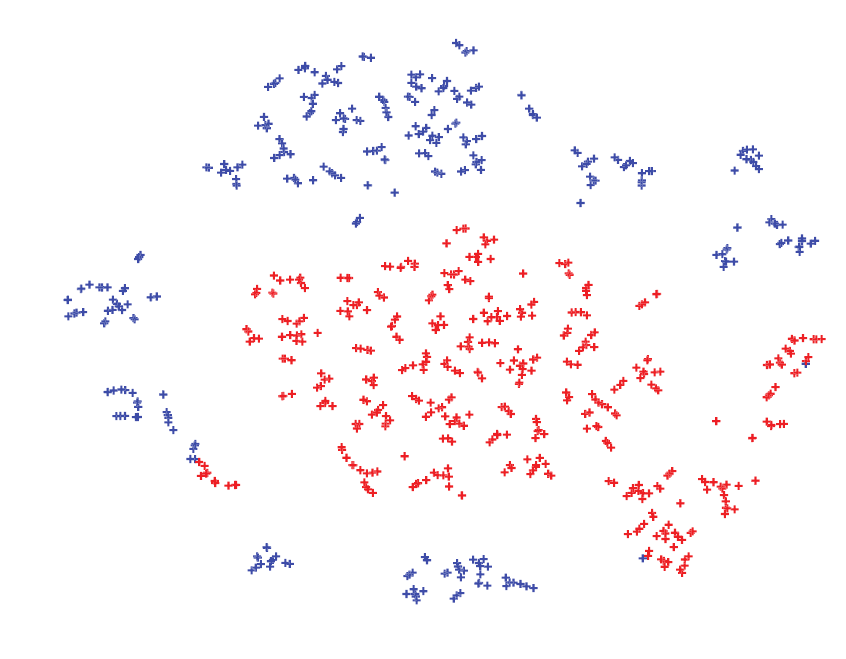}
		\vspace{-15pt}
		\caption{DeepWalk}
	\end{subfigure} \hspace{-7pt}
	\begin{subfigure}[b]{0.16\textwidth}
		\centering
		\includegraphics[width=\textwidth]{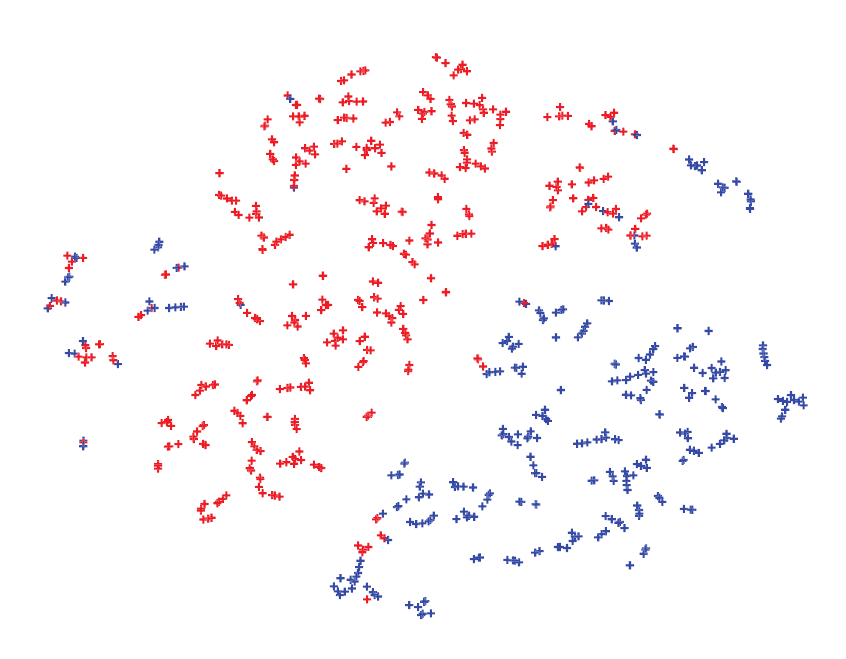}
		\vspace{-15pt}
		\caption{Node2vec}
	\end{subfigure} \hspace{-7pt}
	\begin{subfigure}[b]{0.16\textwidth}
		\centering
		\includegraphics[width=\textwidth]{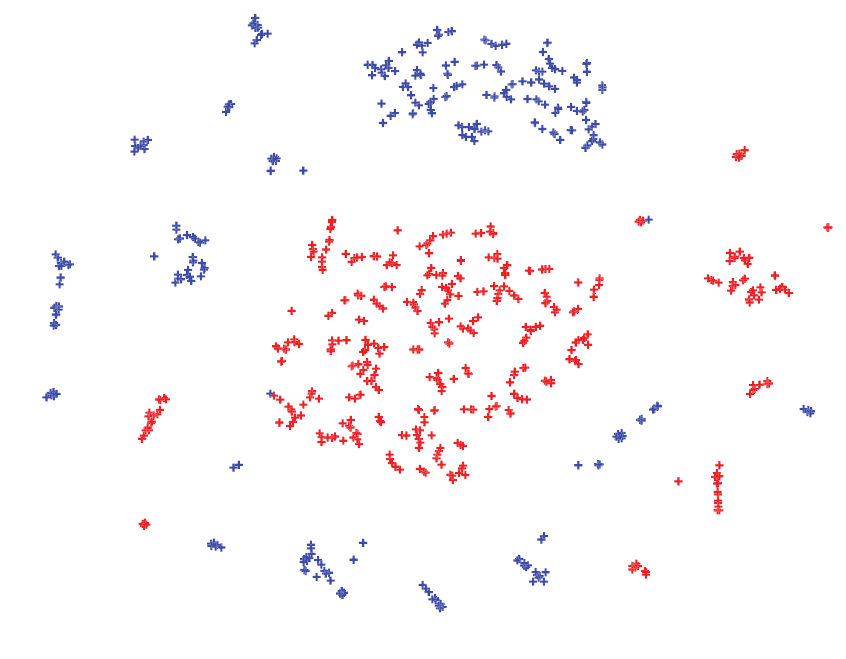}
		\vspace{-15pt}
		\caption{LINE}
	\end{subfigure} \hspace{-7pt}
	\begin{subfigure}[b]{0.16\textwidth}
		\centering
		\includegraphics[width=\textwidth]{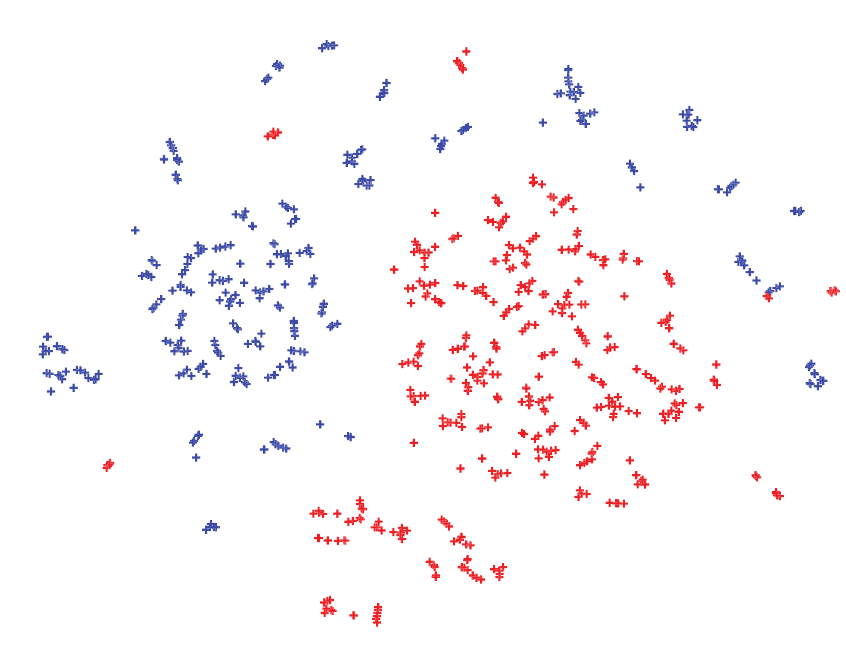}
		\vspace{-15pt}
		\caption{Metapath2vec++}
	\end{subfigure} \hspace{-7pt}
	\begin{subfigure}[b]{0.16\textwidth}
		\centering
		\includegraphics[width=\textwidth]{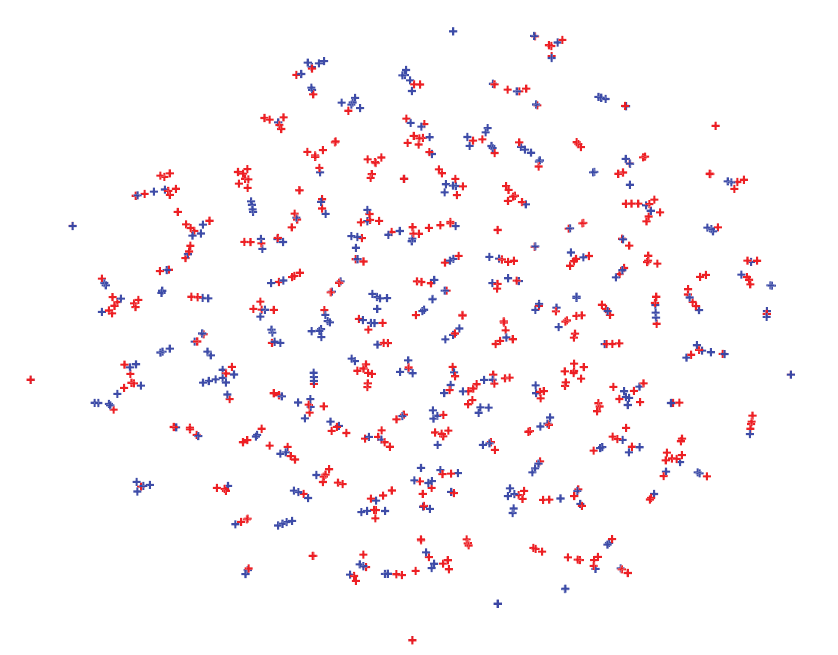}
		\vspace{-15pt}
		\caption{BiNE'}
	\end{subfigure} \hspace{-7pt}
	\begin{subfigure}[b]{0.16\textwidth}
		\centering
		\includegraphics[width=\textwidth]{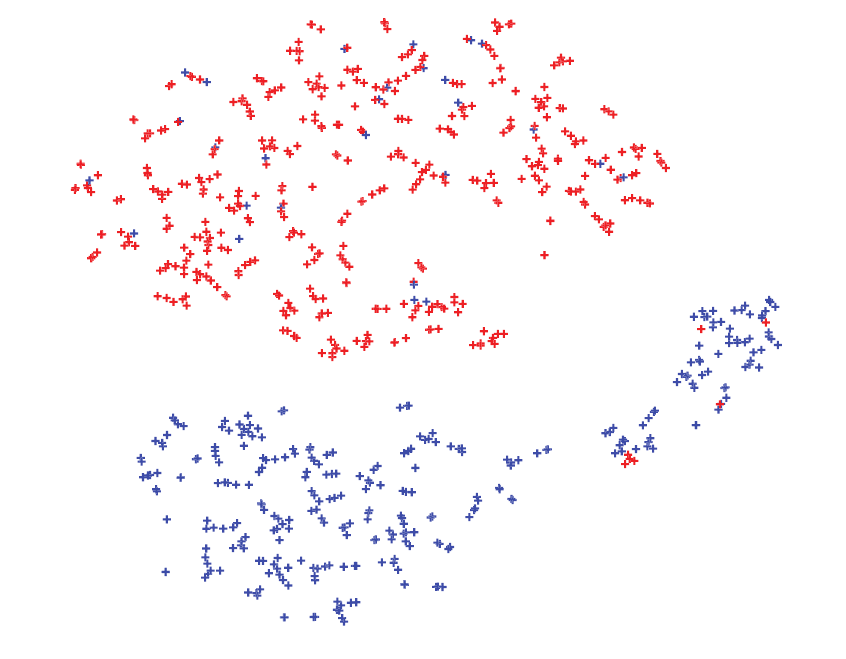}
		\vspace{-15pt}
		\caption{BiNE}
	\end{subfigure} \hspace{-7pt}
	\begin{subfigure}[b]{0.16\textwidth}
		\centering
		\includegraphics[width=\textwidth]{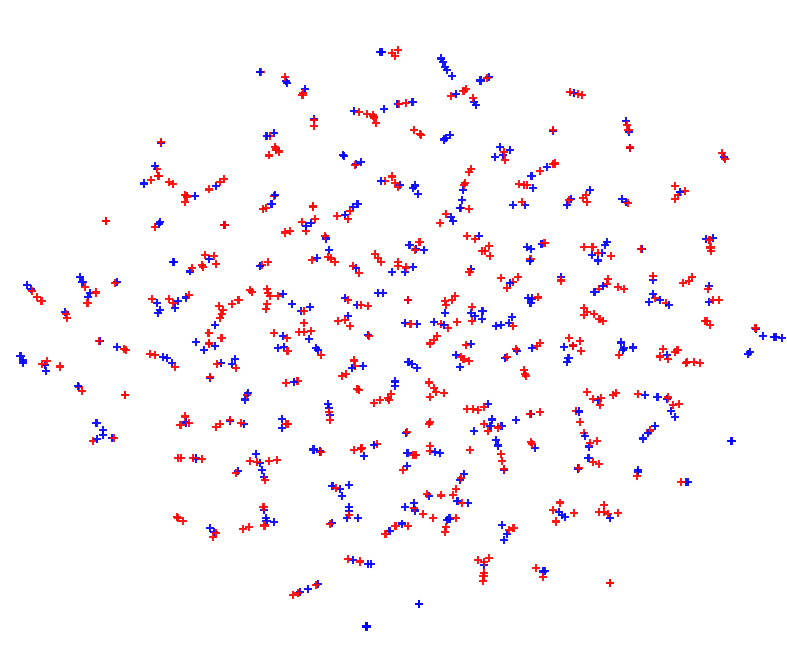}
		\vspace{-15pt}
		\caption{SMF'}
	\end{subfigure} \hspace{-7pt}
	\begin{subfigure}[b]{0.16\textwidth}
		\centering
		\includegraphics[width=\textwidth]{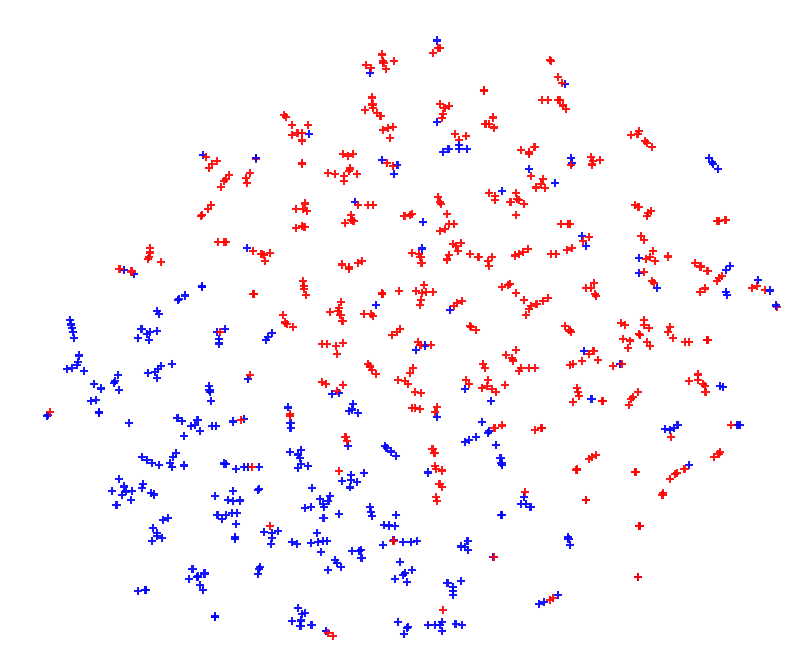}
		\vspace{-15pt}
		\caption{SMF}
	\end{subfigure} \hspace{-7pt}
	\begin{subfigure}[b]{0.16\textwidth}
		\centering
		\includegraphics[width=\textwidth]{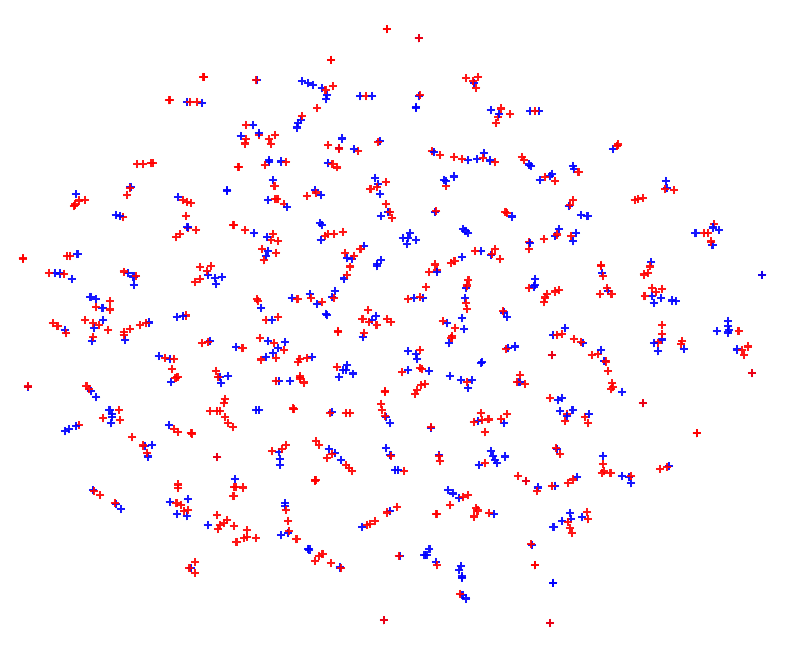}
		\vspace{-15pt}
		\caption{BiNE-MF'}
	\end{subfigure} \hspace{-7pt}
	\begin{subfigure}[b]{0.16\textwidth}
		\centering
		\includegraphics[width=\textwidth]{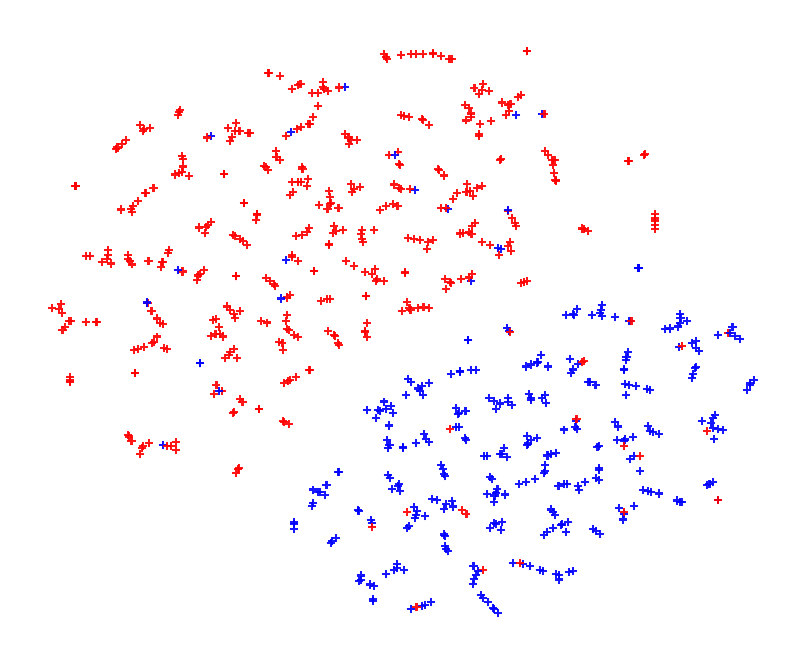}
		\vspace{-15pt}
		\caption{BiNE-MF}
	\end{subfigure} \hspace{-7pt}
	\caption{Visualization of authors in DBLP. Color of a vertex indicates the research fields of the authors (red: ``computer science theory'', blue: ``artificial intelligence''). BiNE', SMF' and BiNE-MF' is the version of BiNE, SMF and BiNE-MF -- without implicit relations.}
	\label{fig:visualization}
\end{figure}
%
We extra a small collaboration bipartite network from DBLP dataset, which contains $736$ researchers and $6$ international journals. A link will be established if the author has published at least $5$ papers on the journal. 
The $6$ journals are from two different research fields: SICOMP, IANDC and TIT from computer science theory, and AI, IJCV,and JMLR from artificial intelligence. As such, from the published venues of the researches, we can inference the research field of them.

We utilize the t-SNE tool~\cite{Laurens2008Visualizing} to map the embedding vectors of authors into 2D space.
In Fig.~\ref{fig:visualization}, we use different color to describe different research fields of the researchers, \ie red: ``computer science theory'', blue: ``artificial intelligence'', and show the visualization results given by different embedding approaches.
From it, we can observe that DeepWalk, Node2vec, LINE, Metapath2vec++, and our BiNE are good since researchers belonging different research fields are well seperated.
As far as we are concerned, BiNE gives a better result due to it generates an obvious gap between two research fields.

Specifically, The MF-based methods -- SMF, BiNE-MF can also provide meaningful visualization. 
In addition, BiNE-MF achieves comparable or even better visualizations than BiNE: not only is there a significant gap between users in different research domains, but users in the same domain are more compact, which verifies that the MF-based methods can achieve more sufficient implicit relationships than BiNE.

However, the variant of BiNE, SMF and BiNE-MF ignoring the implicit relations -- BiNE', SMF' and BiNE-MF' show worse layouts than expected, which illustrate the effective of modeling high-order implicit relations for embedding bipartite networks.

\section{Related Work}
\label{sec:related work}
\subsection{Network Representation Learning}
Our work is related to the neural network-based NRL methods. We first review them from the perspective of network types.

The pioneer work DeepWalk~\cite{DBLP:conf/kdd/PerozziAS14} and Node2vec~\cite{DBLP:conf/kdd/GroverL16} extend the idea of Skip-gram~\cite{DBLP:conf/nips/MikolovSCCD13} to model homogeneous network, which is convert to a corpus of vertex sequences by performing truncated random walks. However, they may not be effective to preserve both explicit and implicit relations of the network.
There are some follow-up works exploiting both 1st-order and 2nd-order proximities between vertices to embed homogeneous networks.
Specifically, LINE~\cite{DBLP:conf/www/TangQWZYM15} learns two separated embeddings for 1st-order and 2nd-order relations; SDNE~\cite{DBLP:conf/kdd/WangC016} and DVNE~\cite{DBLP:conf/kdd/ZhuCW018} incorporates both 1st-order and 2nd-order proximities to preserve the network structure; GraRep~\cite{DBLP:conf/cikm/CaoLX15}, AROPE~\cite{DBLP:conf/kdd/ZhangCWPY018} and NEU~\cite{DBLP:conf/ijcai/YangSLT17} further extends the method to capture higher-order proximities, where NEU efficiently approximate the high-order proximities to learn network representation.
Besides capturing high-order proximities,
there are several proposals to incorporate side information into vertex embedding learning, such as vertex labels~\cite{DBLP:conf/nips/xiaoJH17,DBLP:conf/cikm/LiDHTCL17}, community information~\cite{DBLP:conf/cikm/ChenZH16}, textual content~\cite{Wang:2017:SIGIR}, user profiles~\cite{DBLP:journals/corr/LiaoHZC17}, location information~\cite{DBLP:conf/cikm/XieYWXCW16}, among others.
However, since these methods design for homogeneous networks, they might be suboptimal for learning vertex representations for a bipartite network by ignoring the vertex type information. 
In addition, the ``corpus'' generated by the truncated random walks may not capture the characteristics of the network structure, such as the power-law distribution of vertex degrees.

Metapath2vec++~\cite{dong2017metapath2vec}, HNE~\cite{DBLP:conf/kdd/ChangHTQAH15}, EOE~\cite{DBLP:conf/wsdm/XuWCY17}, and HERec~\cite{DBLP:journals/tkde/ShiHZY19}
are representative vertex embedding methods for heterogeneous networks. Although they can be applied to bipartite network which can be seen as a special type of heterogeneous networks, they are not tailored for learning on bipartite networks.
Specifically, HNE aims to integrate content and linkage structures into the embedding process, and Metapath2vec++ ignores the strength of the relations between vertices and treats the explicit and implicit relations as equally. 
For HERec, the meta-paths in a bipartite network only capture the user-item-user and item-user-item relationships, which are equivalent to only consider the 2nd-order proximity, rather than the high-order proximity.
As such, they are suboptimal for vertex representation learning for a bipartite network.

It is noteworthy that recent works have shown an increase of interest in generating vertex embedding by neighborhood aggregation encoders~\cite{DBLP:journals/corr/KipfW16,DBLP:conf/esws/SchlichtkrullKB18,DBLP:conf/kdd/YingHCEHL18}. However, most of these methods~\cite{DBLP:journals/corr/KipfW16,DBLP:conf/esws/SchlichtkrullKB18} rely on vertex features or attributes.

\subsection{Bipartite Network Modeling}
%
As a ubiquitous data structure, bipartite networks have been mined for many applications, among which vertex ranking is an active research problem.
For example, HITS~\cite{DBLP:conf/soda/Kleinberg98} learns to rank vertices by capturing some semantic relations within a bipartite network.
%
Co-HITS~\cite{DBLP:conf/kdd/DengLK09} incorporates content information of vertices and the constraints on relevance into vertex ranking of bipartite network.
BiRank~\cite{DBLP:journals/tkde/HeGKW17} ranks vertices by taking into account both the network structure and prior knowledge.
LambdaFM~\cite{DBLP:conf/cikm/YuanGJCYZ16} is a ranking based factorization machine by considering different positions in the ranking list have different effects on the performance.
DNS~\cite{DBLP:conf/sigir/ZhangCWY13} also improves the performance of top-N recommendation by designing a series of sampling strategies to sample negative items.

Distributed vertex representation is an alternative way to leverage signals from bipartite network.
Unlike the ranking task, it learns a low dimensional representation of a vertex, which can be seen as the ``features'' of the vertex that preserves more information rather than simply a ranking score.
Latent factor model (LFM),
which has been widely investigated in the field of recommender systems and semantic analysis, is the most representative model.
And a typical implementation of LFM is based on matrix factorization~\cite{DCF,DBLP:conf/uai/RendleFGS09,DBLP:conf/sigir/HeZKC16}.
Recent advances utilize deep learning methods to learn vertex embeddings on the user-item network for recommendation~\cite{DBLP:conf/www/HeLZNHC17}.
It is worth pointing out that these methods are tailored for the recommendation task, rather than for learning informative vertex embeddings.
Moreover, they model the explicit relations in bipartite network only, which can be improved by incorporating implicit relations as shown in~\cite{DBLP:conf/AAAI/LuYu,DBLP:conf/aaai/JiangCYXY16}.

%
%

\section{Conclusions}
\label{sec:conclusion}
We have presented BiNE, a novel approach for  embedding bipartite networks.
It jointly models both the explicit relations and high-order implicit relations in learning the representation for vertices.
Our theoretical result reveals that BiNE can be transferred into the algorithm BiNE-MF, which is a implicit multiple matrix factorization, in a closed form. As a result, it broadens the theoretical understanding of BiNE.
Extensive experiments on several tasks of link prediction, recommendation, and visualization demonstrate the effectiveness and rationality of our BiNE method.

In this work, we have only considered the information revealed in observed edges,
thus it may fail for vertices that have few or even no edges.
Since missing data is a common situation in real-world applications, the observed edges may not contain sufficient signal on vertex relations.
%
To address this issue, we plan to extend our BiNE method to model auxiliary side information, such as numerical features
, textual descriptions
, and among other attributes~\cite{DBLP:journals/corr/LiaoHZC17}.
%
In addition, the bipartite networks in many practical applications are dynamically updated~\cite{DBLP:conf/sigir/HeZKC16}.
%
Thus, we plan to investigate how to efficiently refresh embeddings for dynamic bipartite networks.
%
%

\section{Acknowledgments}
\label{sec:acknowledge}
This work has been supported by the National Key Research and Development Program of China under grant 2016YFB1000905,
and the National Natural Science Foundation of China under Grant No. U1811264,
U1911203,
61877018,
61672234,
61672384,
41775008, 61702275, 
and the Shanghai Agriculture Applied Technology Development Program, China (Grant No.T20170303).

\bibliographystyle{unsrt}
\bibliographystyle{abbrv}
\bibliography{sigproc_simple}

\begin{thebibliography}{10}

\bibitem{DBLP:conf/sigir/0001C0Z18}
Ming Gao, Leihui Chen, Xiangnan He, and Aoying Zhou.
\newblock Bine: Bipartite network embedding.
\newblock In {\em {SIGIR}}, pages 715--724, 2018.

\bibitem{silkroad}
Xiang Wang, Xiangnan He, Liqiang Nie, and Tat-Seng Chua.
\newblock Item silk road: Recommending items from information domains to social
  users.
\newblock In {\em SIGIR}, pages 185--194, 2017.

\bibitem{DBLP:acl/CaoHJCL17}
Yixin Cao, Lifu Huang, Heng Ji, Xu~Chen, and Juanzi Li.
\newblock Bridge text and knowledge by learning multi-prototype entity mention
  embedding.
\newblock In {\em ACL}, pages 1623--1633, 2017.

\bibitem{TriRank}
Xiangnan He, Tao Chen, Min-Yen Kan, and Xiao Chen.
\newblock Trirank: Review-aware explainable recommendation by modeling aspects.
\newblock In {\em CIKM}, pages 1661--1670, 2015.

\bibitem{POH}
Fuli Feng, Xiangnan He, Yiqun Liu, Liqiang Nie, and Tat-Seng Chua.
\newblock Learning on partial-order hypergraphs.
\newblock In {\em WWW}, pages 1523--1532, 2018.

\bibitem{DBLP:journals/corr/abs-1711-08752}
Peng Cui, Xiao Wang, Jian Pei, and Wenwu Zhu.
\newblock A survey on network embedding.
\newblock {\em TKDE}, 2018.

\bibitem{DBLP:conf/kdd/ZhuCW018}
Dingyuan Zhu, Peng Cui, Daixin Wang, and Wenwu Zhu.
\newblock Deep variational network embedding in wasserstein space.
\newblock In {\em {SIGKDD}}, pages 2827--2836, 2018.

\bibitem{DBLP:conf/kdd/PerozziAS14}
Bryan Perozzi, Rami Al{-}Rfou, and Steven Skiena.
\newblock Deepwalk: online learning of social representations.
\newblock In {\em {KDD}}, pages 701--710, 2014.

\bibitem{DBLP:journals/corr/LiaoHZC17}
Lizi Liao, Xiangnan He, Hanwang Zhang, and Tat{-}Seng Chua.
\newblock Attributed social network embedding.
\newblock {\em {TKDE}}, 2018.

\bibitem{DBLP:conf/cikm/LiDHTCL17}
Jundong Li, Harsh Dani, Xia Hu, Jiliang Tang, Yi~Chang, and Huan Liu.
\newblock Attributed network embedding for learning in a dynamic environment.
\newblock In {\em {CIKM}}, pages 387--396, 2017.

\bibitem{dong2017metapath2vec}
Yuxiao Dong, Nitesh~V Chawla, and Ananthram Swami.
\newblock metapath2vec: Scalable representation learning for heterogeneous
  networks.
\newblock In {\em {KDD}}, pages 135--144. ACM, 2017.

\bibitem{DBLP:conf/kdd/ChangHTQAH15}
Shiyu Chang, Wei Han, Jiliang Tang, Guo{-}Jun Qi, Charu~C. Aggarwal, and
  Thomas~S. Huang.
\newblock Heterogeneous network embedding via deep architectures.
\newblock In {\em KDD}, pages 119--128, 2015.

\bibitem{DBLP:conf/wsdm/XuWCY17}
Linchuan Xu, Xiaokai Wei, Jiannong Cao, and Philip~S. Yu.
\newblock Embedding of embedding {(EOE):} joint embedding for coupled
  heterogeneous networks.
\newblock In {\em {WSDM}}, pages 741--749, 2017.

\bibitem{DBLP:conf/kdd/GroverL16}
Aditya Grover and Jure Leskovec.
\newblock node2vec: Scalable feature learning for networks.
\newblock In {\em {KDD}}, pages 855--864, 2016.

\bibitem{DBLP:conf/nips/MikolovSCCD13}
Tomas Mikolov, Ilya Sutskever, Kai Chen, Gregory~S. Corrado, and Jeffrey Dean.
\newblock Distributed representations of words and phrases and their
  compositionality.
\newblock In {\em {NIPS}}, pages 3111--3119, 2013.

\bibitem{DBLP:journals/tkde/HeGKW17}
Xiangnan He, Ming Gao, Min{-}Yen Kan, and Dingxian Wang.
\newblock Birank: Towards ranking on bipartite graphs.
\newblock {\em {TKDE}}, 29(1):57--71, 2017.

\bibitem{pongnumkul2018bipartite}
Suchit Pongnumkul and Kazuyuki Motohashi.
\newblock A bipartite fitness model for online music streaming services.
\newblock {\em Physica A: Statistical Mechanics and its Applications},
  490:1125--1137, 2018.

\bibitem{DBLP:conf/www/TangQWZYM15}
Jian Tang, Meng Qu, Mingzhe Wang, Ming Zhang, Jun Yan, and Qiaozhu Mei.
\newblock {LINE:} large-scale information network embedding.
\newblock In {\em {WWW}}, pages 1067--1077, 2015.

\bibitem{DBLP:conf/AAAI/LuYu}
Lu~Yu, Chuxu Zhang, Shichao Pei, Guolei Sun, and Xiangliang~Zhang and.
\newblock Walkranker: A unified pairwise ranking model with multiple relations
  for item recommendation.
\newblock In {\em {AAAI}}, 2018.

\bibitem{DBLP:conf/aaai/JiangCYXY16}
Meng Jiang, Peng Cui, Nicholas~Jing Yuan, Xing Xie, and Shiqiang Yang.
\newblock Little is much: Bridging cross-platform behaviors through overlapped
  crowds.
\newblock In {\em {AAAI}}, pages 13--19, 2016.

\bibitem{DBLP:conf/complenet/AlzahraniHB14}
Taher Alzahrani, Kathy~J. Horadam, and Serdar Boztas.
\newblock Community detection in bipartite networks using random walks.
\newblock In {\em {CompleNet}}, pages 157--165, 2014.

\bibitem{DBLP:conf/kdd/DengLK09}
Hongbo Deng, Michael~R. Lyu, and Irwin King.
\newblock A generalized co-hits algorithm and its application to bipartite
  graphs.
\newblock In {\em {KDD}}, pages 239--248, 2009.

\bibitem{DBLP:journals/jmlr/LeskovecCKFG10}
Jure Leskovec, Deepayan Chakrabarti, Jon~M. Kleinberg, Christos Faloutsos, and
  Zoubin Ghahramani.
\newblock Kronecker graphs: An approach to modeling networks.
\newblock {\em {JMLR}}, 11:985--1042, 2010.

\bibitem{DBLP:conf/soda/Kleinberg98}
Jon~M. Kleinberg.
\newblock Authoritative sources in a hyperlinked environment.
\newblock In {\em {ACM-SIAM}}, pages 668--677, 1998.

\bibitem{DBLP:conf/icde/HongzhiYin}
Hongzhi Yin, Lei Zou, Quoc Viet~Hung Nguyen, Zi~Huang, and Xiaofang Zhou.
\newblock Joint event-partner recommendation in event-based social networks.
\newblock In {\em ICDE}, 2018.

\bibitem{DBLP:conf/cikm/WangCSR13}
Hongya Wang, Jiao Cao, LihChyun Shu, and Davood Rafiei.
\newblock Locality sensitive hashing revisited: filling the gap between theory
  and algorithm analysis.
\newblock In {\em {CIKM}}, pages 1969--1978, 2013.

\bibitem{DBLP:conf/wsdm/QiuDMLWT18}
Jiezhong Qiu, Yuxiao Dong, Hao Ma, Jian Li, Kuansan Wang, and Jie Tang.
\newblock Network embedding as matrix factorization: Unifying deepwalk, line,
  pte, and node2vec.
\newblock In {\em {WSDM}}, pages 459--467, 2018.

\bibitem{DBLP:conf/www/HeLZNHC17}
Xiangnan He, Lizi Liao, Hanwang Zhang, Liqiang Nie, Xia Hu, and Tat{-}Seng
  Chua.
\newblock Neural collaborative filtering.
\newblock In {\em {WWW}}, pages 173--182, 2017.

\bibitem{DBLP:conf/nips/LevyG14}
Omer Levy and Yoav Goldberg.
\newblock Neural word embedding as implicit matrix factorization.
\newblock In {\em {NIPS}}, pages 2177--2185, 2014.

\bibitem{DBLP:journals/computer/KorenBV09}
Yehuda Koren, Robert~M. Bell, and Chris Volinsky.
\newblock Matrix factorization techniques for recommender systems.
\newblock {\em {IEEE} Computer}, 42(8):30--37, 2009.

\bibitem{DBLP:conf/asunam/XiaDLZX12}
Shuang Xia, Bing~Tian Dai, Ee{-}Peng Lim, Yong Zhang, and Chunxiao Xing.
\newblock Link prediction for bipartite social networks: The role of structural
  holes.
\newblock In {\em {ASONAM}}, pages 153--157, 2012.

\bibitem{DBLP:conf/uai/RendleFGS09}
Steffen Rendle, Christoph Freudenthaler, Zeno Gantner, and Lars
  Schmidt{-}Thieme.
\newblock {BPR:} bayesian personalized ranking from implicit feedback.
\newblock In {\em {UAI}}, pages 452--461, 2009.

\bibitem{DBLP:conf/recsys/TakacsT12}
G{\'{a}}bor Tak{\'{a}}cs and Domonkos Tikk.
\newblock Alternating least squares for personalized ranking.
\newblock In {\em RecSys}, pages 83--90, 2012.

\bibitem{DBLP:conf/kdd/KabburNK13}
Santosh Kabbur, Xia Ning, and George Karypis.
\newblock {FISM:} factored item similarity models for top-n recommender
  systems.
\newblock In {\em {KDD}}, pages 659--667, 2013.

\bibitem{DBLP:conf/sigir/WangYZGXWZZ17}
Jun Wang, Lantao Yu, Weinan Zhang, Yu~Gong, Yinghui Xu, Benyou Wang, Peng
  Zhang, and Dell Zhang.
\newblock {IRGAN:} {A} minimax game for unifying generative and discriminative
  information retrieval models.
\newblock In {\em Proceedings of the 40th International {ACM} {SIGIR}
  Conference on Research and Development in Information Retrieval, Shinjuku,
  Tokyo, Japan, August 7-11, 2017}, pages 515--524, 2017.

\bibitem{Laurens2008Visualizing}
Van Der~Maaten Laurens, Geoffrey Hinton, and Geoffrey Hinton, Van Der~Maaten.
\newblock Visualizing data using t-sne.
\newblock {\em {JMLR}}, 9(2605):2579--2605, 2008.

\bibitem{DBLP:conf/kdd/WangC016}
Daixin Wang, Peng Cui, and Wenwu Zhu.
\newblock Structural deep network embedding.
\newblock In {\em {KDD}}, pages 1225--1234, 2016.

\bibitem{DBLP:conf/cikm/CaoLX15}
Shaosheng Cao, Wei Lu, and Qiongkai Xu.
\newblock Grarep: Learning graph representations with global structural
  information.
\newblock In {\em CIKM}, pages 891--900, 2015.

\bibitem{DBLP:conf/kdd/ZhangCWPY018}
Ziwei Zhang, Peng Cui, Xiao Wang, Jian Pei, Xuanrong Yao, and Wenwu Zhu.
\newblock Arbitrary-order proximity preserved network embedding.
\newblock In {\em {SIGKDD}}, pages 2778--2786, 2018.

\bibitem{DBLP:conf/ijcai/YangSLT17}
Cheng Yang, Maosong Sun, Zhiyuan Liu, and Cunchao Tu.
\newblock Fast network embedding enhancement via high order proximity
  approximation.
\newblock In {\em Proceedings of the Twenty-Sixth International Joint
  Conference on Artificial Intelligence, {IJCAI} 2017, Melbourne, Australia,
  August 19-25, 2017}, pages 3894--3900, 2017.

\bibitem{DBLP:conf/nips/xiaoJH17}
Xiao Huang, Jundong Li, and Xia Hu.
\newblock Label informed attributed network embedding.
\newblock In {\em WSDM}, pages 550--558, 2017.

\bibitem{DBLP:conf/cikm/ChenZH16}
Jifan Chen, Qi~Zhang, and Xuanjing Huang.
\newblock Incorporate group information to enhance network embedding.
\newblock In {\em {CIKM}}, pages 1901--1904, 2016.

\bibitem{Wang:2017:SIGIR}
Chuan-Ju Wang, Ting-Hsiang Wang, Hsiu-Wei Yang, Bo-Sin Chang, and Ming-Feng
  Tsai.
\newblock Ice: Item concept embedding via textual information.
\newblock In {\em SIGIR}, pages 85--94, 2017.

\bibitem{DBLP:conf/cikm/XieYWXCW16}
Min Xie, Hongzhi Yin, Hao Wang, Fanjiang Xu, Weitong Chen, and Sen Wang.
\newblock Learning graph-based {POI} embedding for location-based
  recommendation.
\newblock In {\em {CIKM}}, pages 15--24, 2016.

\bibitem{DBLP:journals/tkde/ShiHZY19}
Chuan Shi, Binbin Hu, Wayne~Xin Zhao, and Philip~S. Yu.
\newblock Heterogeneous information network embedding for recommendation.
\newblock {\em {IEEE} Trans. Knowl. Data Eng.}, 31(2):357--370, 2019.

\bibitem{DBLP:journals/corr/KipfW16}
Thomas~N. Kipf and Max Welling.
\newblock Semi-supervised classification with graph convolutional networks.
\newblock In {\em {ICRL}}, 2016.

\bibitem{DBLP:conf/esws/SchlichtkrullKB18}
Michael~Sejr Schlichtkrull, Thomas~N. Kipf, Peter Bloem, Rianne van~den Berg,
  Ivan Titov, and Max Welling.
\newblock Modeling relational data with graph convolutional networks.
\newblock In {\em {ESWC}}, pages 593--607, 2018.

\bibitem{DBLP:conf/kdd/YingHCEHL18}
Rex Ying, Ruining He, Kaifeng Chen, Pong Eksombatchai, William~L. Hamilton, and
  Jure Leskovec.
\newblock Graph convolutional neural networks for web-scale recommender
  systems.
\newblock In {\em {SIGKDD}}, pages 974--983, 2018.

\bibitem{DBLP:conf/cikm/YuanGJCYZ16}
Fajie Yuan, Guibing Guo, Joemon~M. Jose, Long Chen, Haitao Yu, and Weinan
  Zhang.
\newblock Lambdafm: Learning optimal ranking with factorization machines using
  lambda surrogates.
\newblock In {\em Proceedings of the 25th {ACM} International Conference on
  Information and Knowledge Management, {CIKM} 2016, Indianapolis, IN, USA,
  October 24-28, 2016}, pages 227--236, 2016.

\bibitem{DBLP:conf/sigir/ZhangCWY13}
Weinan Zhang, Tianqi Chen, Jun Wang, and Yong Yu.
\newblock Optimizing top-n collaborative filtering via dynamic negative item
  sampling.
\newblock In {\em The 36th International {ACM} {SIGIR} conference on research
  and development in Information Retrieval, {SIGIR} '13, Dublin, Ireland - July
  28 - August 01, 2013}, pages 785--788, 2013.

\bibitem{DCF}
Hanwang Zhang, Fumin Shen, Wei Liu, Xiangnan He, Huanbo Luan, and Tat-Seng
  Chua.
\newblock Discrete collaborative filtering.
\newblock In {\em SIGIR}, pages 325--334, 2016.

\bibitem{DBLP:conf/sigir/HeZKC16}
Xiangnan He, Hanwang Zhang, Min{-}Yen Kan, and Tat{-}Seng Chua.
\newblock Fast matrix factorization for online recommendation with implicit
  feedback.
\newblock In {\em {SIGIR}}, pages 549--558, 2016.

\end{thebibliography}

\begin{IEEEbiography}[{\includegraphics[width=1in,height=1.25in,clip,keepaspectratio]{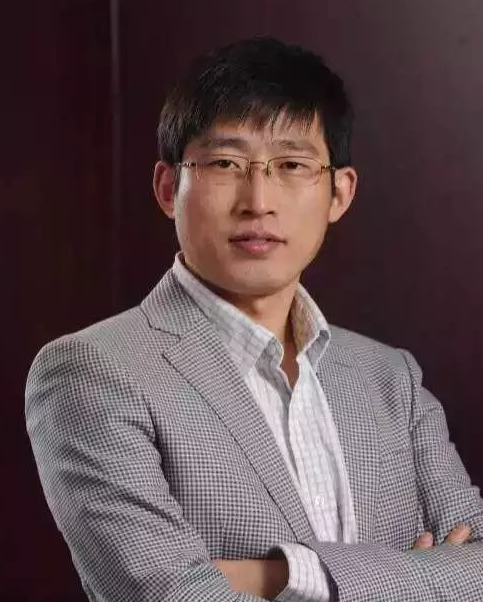}}]{Ming Gao}
	is currently a full professor on School of Data Science and Engineering at East China Normal University, China. He received his doctorate from the School of Computer Science, Fudan University.
%
His research interests include knowledge engineering, user profiling, social network analysis and mining. His works appear in major international journals and conferences, including DMKD, TKDE, KAIS, ICDE, ICDM, SIGIR, etc.
\end{IEEEbiography}
\begin{IEEEbiography}[{\includegraphics[width=1in,height=1.25in,clip,keepaspectratio]{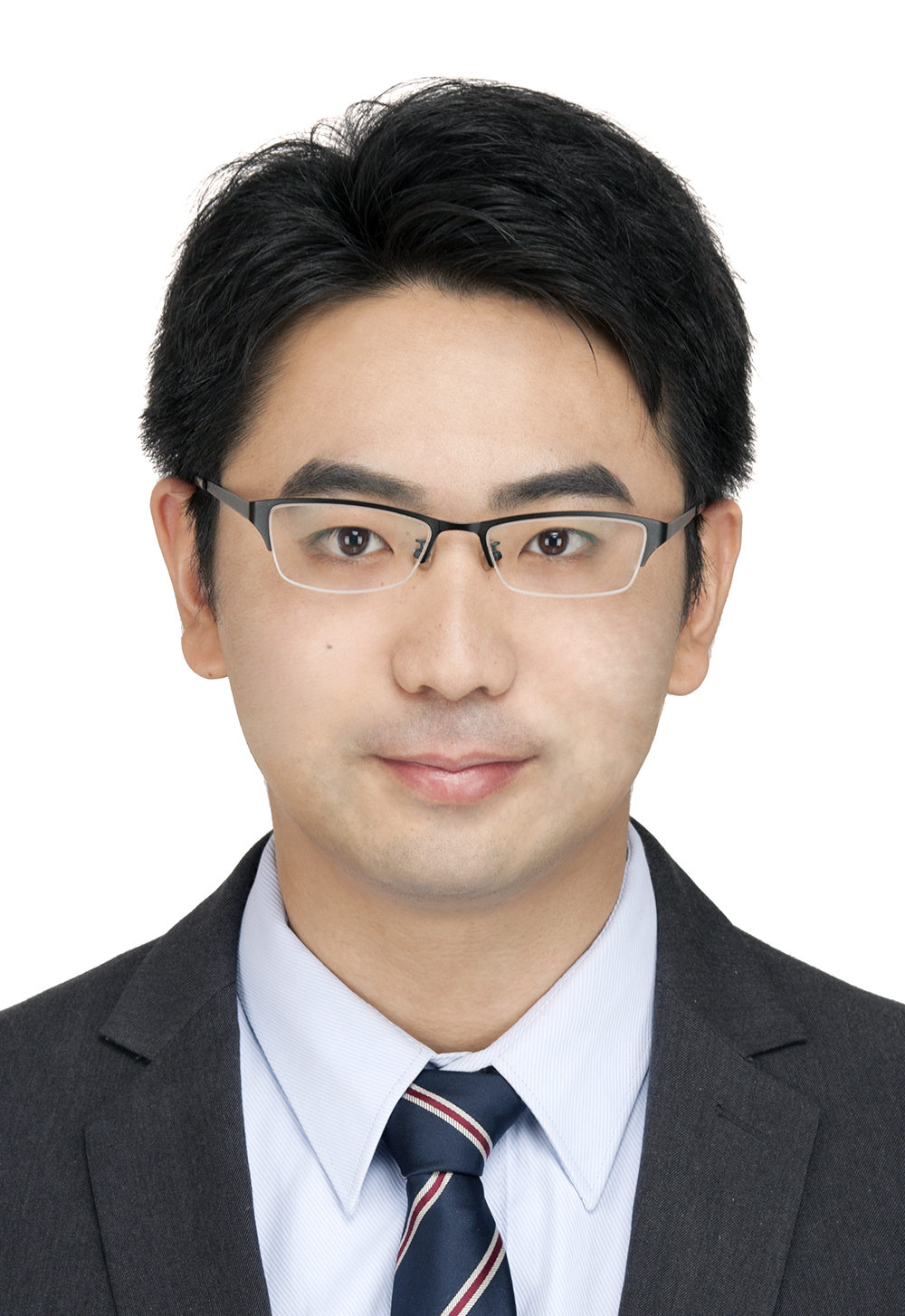}}]{Xiangnan He}
a senior research fellow with School of Computing, National University of Singapore (NUS). He received his Ph.D. in Computer Science from NUS. His research interests span recommender systems, information retrieval, and multi-media processing. He has over 50 publications appeared in several top conferences such as SIGIR, WWW, MM and IJCAI, and journals such as TKDE and TOIS. His work on recommender systems has received the Best Paper Award Honourable Mention of WWW 2018 and SIGIR 2016. Moreover, he has served as the PC member for several top conferences including SIGIR, WWW, MM, KDD etc., and the regular reviewer for journals including TKDE, TOIS, TMM etc.
\end{IEEEbiography} 
\begin{IEEEbiography}[{\includegraphics[width=1in,height=1.25in,clip,keepaspectratio]{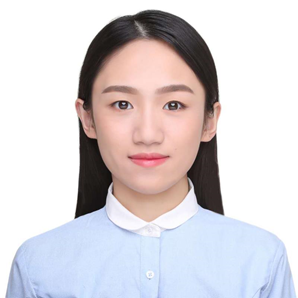}}]{Leihui Chen}
	is currently a master in the School of Data Science and Engineering of East China Normal University, China.
    Her research interests include recommender systems, computational advertising , social network analysis and mining. She is a student member of CCF.
\end{IEEEbiography} 
\begin{IEEEbiography}[{\includegraphics[width=1in,height=1.25in,clip,keepaspectratio]{tingting.png}}]{Tingting Liu}
	is currently a PhD student in the School of Data Science and Engineering of East China Normal University, China.
    Her research interests include recommender systems, knowledge engineering, social network analysis and mining.
\end{IEEEbiography} 
\begin{IEEEbiography}[{\includegraphics[width=1in,height=1.25in,clip,keepaspectratio]{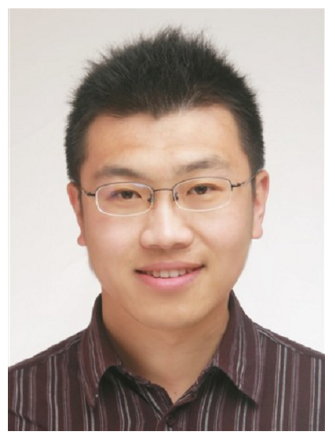}}]{Jinglin Zhang}
	received the MS degree in circuits and systems from Shanghai University, Shanghai, China in 2010 and his PhD degree in electronics and telecommunications from the National Institute of Applied Sciences-Rennes (INSA de Rennes) in 2013. Since 2014, he has been on the faculty of the School of Atmospheric Science, Nanjing University of Information Science Technology.  His current research interest includes computer vision,  high-performance computing, interdisciplinary research with pattern recognition and atmospheric science.
\end{IEEEbiography} 
\begin{IEEEbiography}[{\includegraphics[width=1in,height=1.25in,clip,keepaspectratio]{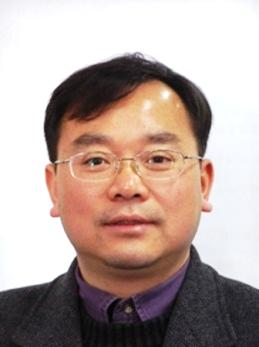}}]{Aoying Zhou}
is a professor on School of Data Science and Engineering (DaSE) at East China Normal University (ECNU), where he is heading DaSE and vice president of ECNU. He is the winner of the National Science Fund for Distinguished Young Scholars supported by NSFC and the professorship appointment under Changjiang Scholars Program of Ministry of Education. His research interests include Web data management, data intensive computing, in-memory cluster computing and benchmark for big data. His works appear in major international journals and conferences, including TKDE, PVLDB, SIGMOD, SIGIR, KDD, WWW, ICDE and ICDM, etc.
\end{IEEEbiography}
\vfill
\end{document}